\documentclass{article}
\usepackage[utf8]{inputenc}
\usepackage{jheppub}
\usepackage{graphicx} 
\usepackage{amsmath,amssymb,commath,braket,ytableau}
\usepackage{cleveref}
\crefname{appsec}{Appendix}{Appendices}

\usepackage{color}
\usepackage{placeins}
\usepackage{tikz}
\usetikzlibrary{arrows.meta,positioning}
\usepackage{makecell}

\usepackage{orcidlink}

\ytableausetup{boxsize=0.4em,aligntableaux=center}

\title{Pseudoreal AMSB: Troubling Tensions with Tumbling}

\author[a,b]{Bea Noether\,\orcidlink{0000-0002-2947-3210}}\emailAdd{bea\_noether@berkeley.edu}

\affiliation[a]{Leinweber Institute for Theoretical Physics, University of California, Berkeley, CA 94720, USA}
\affiliation[b]{Ernest Orlando Lawrence Berkeley National Laboratory, Berkeley, CA 94720, USA}

\abstract{
I analyze the pseudoreal confining $\mathcal{N}=1$ SUSY gauge theories deformed by anomaly mediated SUSY breaking (AMSB). Taking the conjecture that the AMSB and non-SUSY limits are in the same universality class, I find broadly that the non-SUSY confining pseudoreal gauge theories of this class have no massless vectors in their spectra, and that all but one are fully gapped. This contradicts recent expectations from tumbling. I explore whether or not non-standard tumbling could reconcile the two pictures, and find that it would require condensation in repulsive channels in three of the cases, and that in one case it would require condensation in a higher-order channel. I briefly comment on the additional theories that may be in the conformal window, and find that their SUSY versions are consistent with flowing to superconformal fixed points.
}

\begin{document}

\maketitle

\section{Introduction}

In recent years AMSB has proven to be an fruitful method for investigating the IR dynamics of non-SUSY gauge theories. This method has been used to corroborate known results and predict new ones in vectorlike theories like QCD \cite{kondo_near-susy_2025,murayama_exact_2021,kondo_broken_2025}. Particularly interesting has been the application to chiral gauge theories \cite{leedom_exact_2025,csaki_exact_2021,csaki_more_2022,gherghettaHighQualityAxionExact2025}, where in the absence of the lattice results or realizations in Nature enjoyed by vectorlike theories, the only other means for an ansatz to the IR theory comes from the tumbling hypothesis \cite{raby_tumbling_1980}.

An interesting third category of theories, the pseudoreal gauge theories, will be the focus of this work. These are theories with an odd number of Weyl fermions in a pseudoreal representation of the gauge group. Similar to vectorlike theories, pseudoreal theories have a real Euclidean path integral. But they also have no non-trivial mass terms, as in chiral theories. There are only 19 such theories that are asymptotically free \cite{cacciapaglia_selection_2026}. 

Recently, \cite{cacciapaglia_perusing_2026} used tumbling to propose IR dynamics for the pseudoreal gauge theories. They propose that many of these theories should have massless spin-1 particles in their IR spectra. Ten of the theories they considered have SUSY versions that are asymptotically free, of which six\footnote{Notably, these six are the only ones of the 19 considered in \cite{cacciapaglia_perusing_2026} that were determined to be outside the conformal window. Two were listed as indeterminate, and they also have asymptotically free SUSY versions that I analyze in \cref{sec:InCW}.} were conjectured to be outside of the conformal window. As they noted, AMSB can provide a check on these results. 

Already AMSB and tumbling have conflicted in several examples. In \cite{csaki_exact_2021,csaki_more_2022} AMSB was applied to theories of the form $SU(N)+\ydiagram{1,1}+(N-4)\ydiagram{1}$ (``Georgi-Glashow type") and $SU(N)+\ydiagram{2}+(N+4)\overline{\ydiagram{1}}$ (``Bars-Yankielowicz type"), and in both cases results could only be reconciled with tumbling if the NMAC was also allowed to condense. Both those classes of models were also studied with functional RG in \cite{liDynamicalSymmetryBreaking2025} and \cite{liConfinementSymmetryBreaking2026}, respectively, and were found to disagre with the AMSB results. More recently, in \cite{leedom_exact_2025} AMSB was applied to theories of the form $SU(N_c)+\ydiagram{1,1}+N_f\ydiagram{1}+(N_c+N_f-4)\overline{\ydiagram{1}}$ and found ``competitions among various attractive channels." In \cite{kondoDynamicsSimplestChiral2026} AMSB was applied to $SO(10)+N_f\times 16_{\rm spin}$ and found agreement with tumbling for $N_f=1$ and $3$, but disagreement when $N_f=2$. In \cite{gohDynamicsE6Chiral2025} AMSB was applied to $E_6+N_f\times 27_{F}$ and found agreement with tumbling for $N_f=1$ but disagreement for $N_f=2$ and $3$. In many of these cases the predictions regarding the massless spectra disagree starkly.

The IR spectrum of AMSB theories is often exactly calculable. Global and gauge symmetries, as well as massless degrees of freedom, can in such cases be unambiguously determined. Taking seriously the conjecture that the AMSB and non-SUSY theories belong to the same universality class, one can make definitive conclusions about the properties of the non-SUSY theory. Conversely, disagreement with known properties of the non-SUSY limit would establish a given case as a counterexample to the conjecture. 

In this work, I analyze the asymptotically free pseudoreal $\mathcal{N}=1$ SUSY theories perturbed by AMSB. I primarily focus my attention to the six theories whose non-SUSY limits correspond to those determined to be non-conformal in \cite{cacciapaglia_perusing_2026}. In each case, I find that the AMSB theories have no $U(1)$ gauge factor that is not part of a larger non-abelian factor, and thus no massless spin-1 particles in the spectra. This is in tension with the conclusion of \cite{cacciapaglia_perusing_2026}. Since those conclusions, arrived at through the tumbling hypothesis, are also conjecture, this tension cannot immediately be resolved one way or the other. Moreover, the AMSB result would imply condensation in \textit{repulsive} tumbling channels in some cases, which is a much starker disagreement than in any previous cases. In one case, no tumbling realization via bilinear condensates is possible at all. Non-perturbative results from first principles for the non-SUSY theories, such as lattice simulations, are required.

I also briefly discuss the four theories whose non-SUSY limits may be in the conformal window. These SUSY theories do not appear to be discussed in the literature, so I initiate that analysis. I find that they are consistent with flowing to interacting SCFTs in the IR. This does not necessarily mean the non-SUSY limits are conformal, as AMSB could deflect the RG flow away from the fixed point as in \cite{kondo_broken_2025}. The AMSB vacua are not calculable without a detailed understanding of the approach to the fixed point, so I am unable to make any conclusions for these four theories.

Throughout I use the convention that $Sp(2N)$ refers to the symplectic group whose fundamental representation is $2N$-dimensional. I use $\text{Spin}(N)$ to refer to the double cover of $SO(N)$, who both have the same Lie algebra. 

The paper is organized as follows. In \cref{sec:BasicArgument} I provide the simple reason that the AMSB theories do not have massless spin-1 particles in the IR. In \cref{sec:AMSBResults} I review the form of the AMSB deformation, and describe the vacua of each of the AMSB theories. In \cref{sec:InCW} I briefly discuss the pseudoreal theories that may be in the conformal window. In \cref{sec:ReconcileTumbling} I describe how the AMSB symmetry breaking patterns can only be achieved through condensates in repulsive channels. I conclude in \cref{sec:Conclusion}.

\begin{table}
    \centering
    \begin{tabular}{cccc}
        $G_{\rm gauge}$ & Matter & AMSB Subgroup & Tumbling Subgroup\\ \hline \hline
         $SU(6)$& $\ydiagram{1,1,1}$ & $SU(3)\times SU(3)$ & \makecell{$SU(5)\times U(1)$ \\ $SU(4)\times SU(2)\times U(1)$ \\ $SU(3)\times SU(3)\times U(1)$} \\\hline 
         $E_7$& $\ydiagram{1}$ &$E_6$  & $E_6\times U(1)$\\ \hline
         $\text{Spin}(11)$& $32_{\rm spin}$ & $SU(5)$ & \makecell{$\text{Spin}(10)$ \\ $\text{Spin}(9)\times U(1)$}\\ \hline
         $\text{Spin}(12)$& $32_{\rm spin}$ & $SU(6)$ & \makecell{$SU(6)\times U(1)$ \\ $\text{Spin}(10)\times U(1)$}\\\hline
         $\text{Spin}(13)$& $64_{\rm spin}$ & $SU(3)\times SU(3)$ & \makecell{$\text{Spin}(11)\times U(1)$ \\ $\text{Spin}(10)\times SU(2)$}\\\hline
         $Sp(6)$& $\ydiagram{1}+\ydiagram{1,1,1}$ & $SU(2)$  & \makecell{$SU(3)\times U(1)$ \\ $SU(2)\times Sp(4)$ \\ $SU(2)\times U(1)\times U(1)$}\\
    \end{tabular}
    \caption{The six pseudoreal theories with asymptotically free $\mathcal{N}=1$ versions that were found in \cite{cacciapaglia_perusing_2026} to be outside the conformal window. The columns show the gauge group $G_{\rm gauge}$, the matter representations, the stabilizer subgroup in the AMSB theory, and the stabilizer subgroup(s) in the tumbling hypothesis.}
    \label{tab:mainresults}
\end{table}

\section{Basic Argument}\label{sec:BasicArgument}

The key input is that, in a 4d gauge theory, a massless spin-1 state can only arise from an
\emph{unbroken gauge symmetry} in the vacuum. In an AMSB vacuum which lies (parametrically) near the
supersymmetric $D$-flat directions, the gauge symmetry breaking pattern is therefore controlled by the
stabilizer of the corresponding $D$-flat point: broken generators acquire masses via the Higgs mechanism,
while unbroken generators may remain light at the perturbative level.

However, the only way an unbroken generator can yield a genuinely massless vector in the deep IR is
if it corresponds to an \emph{isolated} abelian factor of the residual gauge group. By ``isolated''
we mean a $U(1)$ factor that is \emph{not embedded} in a remaining non-abelian gauge factor
(i.e. the residual gauge group does not contain a non-abelian factor whose confinement would remove the
corresponding Cartan gauge bosons from the massless spectrum). When the residual gauge symmetry contains
a non-abelian factor with no massless matter, the supersymmetric dynamics confines and produces a mass gap.
Moreover, this persists under anomaly-mediated SUSY breaking: it is known that pure SYM remains gapped
under the AMSB deformation, so the low-energy spectrum cannot contain massless spin-1 states sourced by such
non-abelian factors.

In the six pseudoreal theories analyzed in this work, the supersymmetric limit exhibits ADS-like
runaway behavior. The AMSB deformation stabilizes the would-be runaway at large, nonzero field values,
in a regime where the perturbative description is self-consistent and where the AMSB minimum is well-approximated
by a $D$-flat vacuum. Along every quantum-allowed $D$-flat direction relevant for the AMSB stabilization,
the residual gauge symmetry never contains an isolated $U(1)$ factor: for each AMSB vacuum direction, all
UV gauge bosons are either
(i) Higgsed (and hence become massive vector multiplets), or
(ii) belong to a remaining non-abelian factor that confines in the IR. I summarize this discrepancy between the stabilizers in the AMSB theory and those of the tumbling hypothesis in \cref{tab:mainresults}.

Therefore, using only SUSY gauge symmetry breaking and confinement data, we can already conclude that the AMSB
theories contain no massless spin-1 particles in their spectra. In fact, for the all but one of the cases at hand the same reasoning
implies that the AMSB deformation fully gaps the IR: besides the absence of an isolated $U(1)$ photon,
any remaining $D$-flat moduli are lifted by AMSB, while the confining sectors produce only massive singlet
composites. I explicitly minimize each of the AMSB-deformed theories, and summarize the the residual symmetries and spectra in \cref{sec:AMSBResults}.

\section{The AMSB Results}\label{sec:AMSBResults}

Anomaly mediated SUSY breaking was introduced  in \cite{giudiceGauginoMassSinglets1998,randallOutThisWorld1999}. It can be conveniently encoded in a nonzero F-term for the Weyl compensator $\Phi = 1 + m \theta^2$, where $m$ is the SUSY breaking scale and corresponds to a gravitino mass if one considers the AMSB RG trajectories to be embedded in a higher-dimensional supergravity theory.

There are tree-level SUSY breaking contributions to the scalar potential:
\begin{align}
    V_{\rm tree} =& \partial_i W g^{i\bar{j}} \partial_{\bar{j}}W^* + \abs{m}^2 (\partial_i K g^{i\bar{j}}\partial_{\bar{j}}K^* - K) + m (\partial_i W g^{i\bar{j}}\partial_{\bar{j}}K-3W) + h.c.
    \label{eq:Vtree}
\end{align}
where $W$ is the superpotential, $K$ is the K\"ahler potential, and $g^{i\bar{j}}$ is the inverse of the K\"ahler metric $g_{i\bar{j}} = \partial_i\partial_{\bar{j}}K$. It should be noted that when the superpotential contains no dimensionful couplings, the SUSY-breaking $m$-dependent terms in \cref{eq:Vtree} vanish identically.

There are also loop-level SUSY breaking contributions to the scalar and gaugino masses, as well as the trilinear couplings
\begin{align}
    m_\lambda =& -\frac{\beta(g^2)}{2g^2}m
    &
    m_\phi^2 =& -\frac{1}{4}\dot{\gamma}_i\abs{m}^2
    &
    A_{ijk} =& -\frac{1}{2}(\gamma_i+\gamma_j+\gamma_k)m
    \label{eq:masses+trilinear}
\end{align}
where\footnote{Throughout the literature there is often a difference in convention for the anomalous dimensions by a sign or factor of two. I use this convention, which is prevalent in AMSB literature.} $\gamma_i = \dod{\ln Z_i}{\ln\mu}$ is the anomalous dimension corresponding to the wavefunction renormalization $Z_i$ of the field $i$, $\dot{\gamma}_i = \dod{\gamma_i}{\ln\mu}$, and $\beta(g^2) = \dod{g^2}{\ln\mu}$. We typically take $m>0$ by convention, but it should be noted that the phase of $m$ can be absorbed via a $U(1)_R$ transformation. From the above, it can be seen that when the gauge theory is asymptotically free $\dot{\gamma}_i<0$ and the scalar masses-squared are positive, and thus the vacuum of the AMSB theory is stabilized. When applied as a deformation to an asymptotically free SUSY theory with massless matter, the scalars and gauginos become massive and the dynamics at scales well below $m$ is described by a theory of gauge bosons and massless fermions. 

Crucially, AMSB has the property of UV insensitivity: the form of the soft terms at a given scale depends only on the degrees of freedom at that scale. That is, the SUSY breaking ``commutes" with the integrating out of heavy multiplets, so long as they are above the SUSY breaking scale $m$. This is particularly useful in a context where exact results are known about the strongly-coupled regime of an asymptotically free SUSY theory. The theory can be unambiguously defined at some perturbative scale $\mu_{\rm UV}\gg \Lambda$, and an exact effective description is known at some non-perturbative scale $\mu_{\rm IR}\ll \Lambda$ (the seminal example being \cite{seibergElectricMagneticDuality1995}). The result of applying the AMSB deformation at the scale $\mu_{\rm UV}$ is guaranteed to flow to the result of applying the deformation at the scale $\mu_{\rm IR}$, so long as $m\ll \mu_{\rm IR}$. This is in contrast to most other kinds of SUSY breaking, in which the IR and UV deformations are generically not on the same RG trajectory. 

This motivated the ``AMSB Conjecture" first put forward in \cite{murayama_exact_2021}, that the AMSB-deformed $\mathcal{N}=1$ version of a strongly coupled gauge theory should be in the same vacuum universality class as its non-SUSY $m/\Lambda \to\infty$ limit. This approach has been applied to a wide variety of theories in the past few years (see \cite{csaki_exact_2021,csaki_more_2022,royvarierAMSBSpNcGauge2026,csakiPhaseTransitionsUnusual2025a,gohDynamicsE6Chiral2025,leedom_exact_2025,kondoDynamicsSimplestChiral2026,csakiPhasesNonsupersymmetricGauge2021,csakiDemonstrationConfinementChiral2021,delimaSconfiningSUSYQCDAnomaly2023,baiNovelPhasesBaryondense2025,tearlachrobertsonorkishAnomalyMediationSeibergWitten2026}). In particular, AMSB has been used to reproduce and extend a number of non-trivial results about QCD from first principles \cite{csakiGuideAnomalymediatedSupersymmetrybreaking2023,kondo_broken_2025,kondo_near-susy_2025}.

For an asymptotically free $\mathcal{N}=1$ theory with scale $\Lambda$, we can use exact results about the SUSY theory to work out exact features of the AMSB deformed theory so long as the SUSY breaking is small $m\ll \Lambda$. Conversely, when $m\gg \Lambda$ all of the superpartners decouple at high energies and the only degrees of freedom relevant for dynamics at the scale $\Lambda$ are those of the non-SUSY theory, so in this limit the IR is rather uncontroversially that of the non-SUSY theory. The conjecture is solely about whether or not there is a phase transition in the intermediate regime $m\sim \Lambda$. Explicit examples where such an intermediate-scale phase transition is actually known to occur
are very limited, and are associated with classically conformal dynamical superpotentials
\cite{baiPhasesConfiningSU52022,delimaSconfiningSUSYQCDAnomaly2023}. In cases where no such phase transition occurs, the vacuum structure (massless degrees of freedom, global symmetries, etc.) of the AMSB theory is of the same class as that of the non-SUSY theory of interest.
While a fully rigorous proof is elusive, the Wilsonian/universality picture strongly suggests that no intermediate IR-restructuring transition occurs along the soft-breaking trajectory \cite{dineChallengesObtainingResults2022,dinePossibilityDemonstratingConfinement2022}.

In the remainder of this section, I systematically work out details of the vacua and spectra of the AMSB-deformed pseudoreal $\mathcal{N}=1$ theories. In each case I find stark disagreement with the conclusions of the tumbling analysis done in \cite{cacciapaglia_perusing_2026}. The main results can be summarized via the stabilizer subgroups of the minima, which I list in \cref{tab:mainresults}. In short, I find that there are no massless vectors in the spectra of the AMSB theories. When listing the spectra, I list the states grouped by organization into $\mathcal{N}=1$ multiplets with the understanding that there will be splittings among these multiplets. The precise values of all the masses are exactly calculable in AMSB, and while that would certainly be of interest in any phenomenological application for my purposes the rough scaling is all that is needed.

\subsection{$SU(6)$ with $20_{A3}$}\label{sec:SU6withA3}
It is worth noting that this is the only pseudoreal theory that is ``t-confining" as defined by~\cite{ishikawa_truly_2025}. Let $A$ denote the $20_{A3}$. The D-flat constraint can be written as\footnote{Here $A^*$ denotes the complex-conjugate, not to be confused with the Hodge dual $\star A$.} 
\begin{align}
A_{i jk}\,A^{*\,\ell jk}
\;=\;\frac{1}{6}\,\delta_i{}^{\ell}\,\bigl(A_{mnp}A^{*\,mnp}\bigr)\,.
\end{align}
The orbit of a generic D-flat vev for $A_{ijk}$ can be represented as
\begin{align}
    \braket{A_{123}} = \braket{A_{456}} =& v,
\end{align}
with $v\in\mathbb{C}$. As such D-flat vevs break $SU(6)\to SU(3)\times SU(3)$ with no remaining massless charged matter. There is no residual isolated $U(1)$. The non-abelian gauge factors confine and the AMSB theory cannot have a massless spin-1 particle. 

For completeness, we analyze the AMSB theory explicitly. From \cite{dotti_affine_1998}, the dynamical superpotential generated by gaugino condensation in the two $SU(3)$ factors is
\begin{align}
    W =& \frac{\Lambda^5}{v^2}(\omega^r -\omega^s),
\end{align}
where $r,s = 0,1,2$ denote the different branches and $\omega = e^{2\pi i/3}$. The AMSB potential when $r\neq s$ and thus $W\neq 0$ is then given by
\begin{align}
    V =& 2\frac{\abs{\Lambda}^{10}}{\abs{v}^6}(\omega^r-\omega^s)(\omega^{-r}-\omega^{-s}) -5m\frac{\Lambda^5}{v^2}(\omega^r-\omega^s) + h.c.\, .
\end{align}
Note that $\abs{\omega^r-\omega^s}=\sqrt{3}$ for all $r\neq s$. This is minimized for 
\begin{align}
    v =& \Lambda\left(\frac{3^{3/2}\Lambda}{5m}\right)^{1/4}
    e^{i\frac{1}{2}\arg(\omega^r-\omega^s)},\qquad r\neq s,
\end{align}
with $V_{\rm min} = -\frac{20 \sqrt{5} }{3^{5/4}}\Lambda ^{5/2} m^{3/2}$. Note that there are six such degenerate minima, consistent with the $\mathbb{Z}_6$ center symmetry. 

On the $W=0$ branch, however, we can follow the flat direction down to small field values where the description becomes strongly coupled. In terms of some putative dual IR variable $\mathcal{A}$, the K\"ahler potential in this limit includes higher order terms:
\begin{align}
K =& \mathcal{A}^\dagger\mathcal{A} + \frac{a}{\Lambda^2}(\mathcal{A}^\dagger\mathcal{A})^2 + \cdots     ,
\end{align}
where $a$ is presumably an order-one coefficient. This description is only valid for $\mathcal{A}\lesssim \Lambda$, after which the perturbative description becomes valid and we know the potential vanishes. For $\mathcal{A}\ll \Lambda$ the leading order piece in the tree-level potential is\footnote{The leading order pieces of $\partial_i K g^{i\bar{j}}\partial_{\bar{j}}K^* - K$ cancel, while the next-to-leading-order piece just comes from $K$.} $V_{\rm tree}\sim - a \abs{\mathcal{A}}^4/\Lambda^2 $. At this point we cannot make progress without some information about the behavior of the K\"ahler potential in the regime $\mathcal{A}\sim\Lambda$. If we make the plausible but non-rigorous assumption\footnote{I thank the authors of~\cite{ishikawa_truly_2025} for bringing my attention to this simple argument against the global minimum being in the strongly-coupled region.} that the $\mathcal{A}\sim\Lambda$ interpolates between the small and large limits without introducing any sort of ``sudden drop" we can then estimate the size of a minimum by capping off the perturbative descent at $\mathcal{A}\sim\Lambda$. Under such a regime, the minimum along the $W =0$ branch can be estimated\footnote{This is of course also assuming the coefficient $a$ in the K\"ahler potential is not tuned to be very large for some reason, such as $a\sim (\Lambda/m)^p$ for some $p>0$.} to be $V\sim - m^2\Lambda^2$. The minimum along the $W\neq 0$ branches, however, scale as $V\sim - m^{3/2}\Lambda^{5/2}$. For sufficiently small $m\ll\Lambda$ the $W\neq 0$ minimum is thus deeper, subject to these assumptions on the K\"ahler potential.

I move forward with the assumption that the global minimum lies on one of the $W\neq 0$ branches (so that the Higgsed description and AMSB potential are reliable). The Higgsing scale is parametrically
\begin{align}
m_V\sim g_6\,|v|\sim g_6\,\Lambda\left(\frac{\Lambda}{m}\right)^{1/4}\,.
\end{align}
Under $SU(3)_1\times SU(3)_2$,
\begin{align}
20\rightarrow (1,1)_{S_1}\oplus(1,1)_{S_2}\oplus(\bar 3,3)\oplus(3,\bar 3)\,,
\end{align}
with $\langle S_1\rangle=\langle S_2\rangle=v$ and $\langle(\bar 3,3)\rangle=\langle(3,\bar 3)\rangle=0$. The adjoint decomposes as
\begin{align}
35\rightarrow (8,1)\oplus(1,8)\oplus(3,\bar 3)\oplus(\bar 3,3)\oplus(1,1)\,,
\end{align}
so the broken generators lie in $(3,\bar 3)\oplus(\bar 3,3)\oplus(1,1)$, i.e. $19$ massive $\mathcal N=1$ vector multiplets. The $(\bar 3,3)\oplus(3,\bar 3)$ components of $A$ provide the Goldstone directions for the $18$ complex broken generators, while one linear combination of the two singlets $S_1,S_2$ is eaten; the orthogonal singlet combination survives as the single physical chiral multiplet along the D-flat direction. Degree-of-freedom matching at $m_V$ is consistent.

Below $m_V$, both $SU(3)$ factors reduce to pure $\mathcal N=1$ SYM and confine. Using gaugino condensation, the confinement scale can be read off from the form of the dynamical superpotential above as
\begin{align}
\Lambda_3^3\sim \frac{\Lambda^5}{v^2}
\quad\Rightarrow\quad
\Lambda_3\sim \left(\frac{\Lambda^5}{v^2}\right)^{1/3}\sim \Lambda^{5/6}m^{1/6}\,.
\end{align}
Thus each confining $SU(3)$ sector yields only color-singlet massive glueball/gluinoball composites with characteristic mass $O(\Lambda_3)$, and there are no massless spin-1 states.

The remaining gauge-singlet chiral modulus (the physical D-flat direction orthogonal to the eaten combination) is stabilized by AMSB at $m_{\rm mod}\sim m\,$.

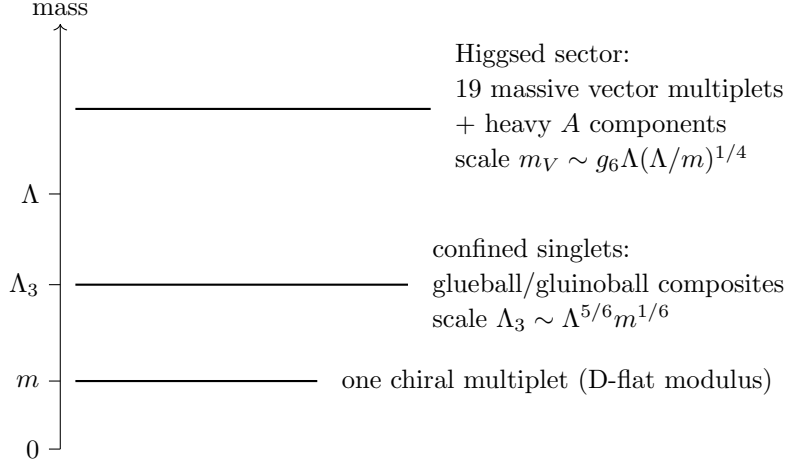
\begin{figure}[t]
\centering
\begin{tikzpicture}[x=1cm,y=0.75cm]
  \draw[->] (0,0) -- (0,7.5) node[above] {mass};

  \draw (0,0)   -- (-0.15,0)   node[left] {$0$};
  \draw (0,1.2) -- (-0.15,1.2) node[left] {$m$};
  \draw (0,4.5) -- (-0.15,4.5) node[left] {$\Lambda$};

  \draw (0,2.9) -- (-0.15,2.9) node[left] {$\Lambda_3$};

  \draw[thick] (0.2,1.2) -- (3.4,1.2);
  \node[right,align=left] at (3.6,1.2)
    {one chiral multiplet\ (D-flat modulus)};

  \draw[thick] (0.2,2.9) -- (4.6,2.9);
  \node[right,align=left] at (4.8,2.9)
    {confined singlets:\\glueball/gluinoball composites\\scale $\Lambda_3\sim \Lambda^{5/6}m^{1/6}$};

  \draw[thick] (0.2,6.0) -- (4.9,6.0);
  \node[right,align=left] at (5.1,6.0)
    {Higgsed sector:\\19 massive vector multiplets\\+ heavy $A$ components\\scale $m_V\sim g_6 \Lambda (\Lambda/m)^{1/4}$};

\end{tikzpicture}
\caption{Parametric mass spectrum in the $W\neq 0$ vacuum for the $SU(6)$ theory with $20_{A3}$. It should be understood that there is splitting among the $\mathcal{N}=1$ multiplets.}
\label{fig:SU6A3spectrum}
\end{figure}

Collecting the parametric spectrum around a $W\neq 0$ minimum: $\,19$ massive vector multiplets with $m_V\sim g_6 v$, one light chiral multiplet with mass $\sim m$, and confined singlet composites from the two pure $SU(3)$ sectors at $\Lambda_3\sim \Lambda^{5/6}m^{1/6}$. There are no massless charged states and no isolated massless gauge bosons. I summarize the parametric form of the spectrum in \cref{fig:SU6A3spectrum}.

\subsection{$E_7$ with $56_F$}
Let $\phi$ denote the fundamental. Parametrize $\phi$ by the maximal subgroup $E_6\times U(1)$, under which it transforms as $56=1_{3}\oplus 27_1\oplus\overline{27}_{-1}\oplus 1_{-3}$. Write this as $\phi = (s,Z,\tilde{Z},t)$. The D-flat vacua can be of four different forms depending on the rank. Let $e_1,e_2$ be orthogonal primitive idempotents in $J_3(\mathbb{O})$ (for instance $e_1 = \text{diag}(1,0,0)$ and $e_2=\text{diag}(0,1,0)$) and let $E = e_1$ and $U=e_1+e_2$. The only gauge invariant is the quartic $\mathcal{I}_4(\phi)$. For a generic rank 4 vev $\mathcal{I}_4(\phi)\neq 0$ and $E_7\to E_6$ with no isolated $U(1)$ and no remaining massless charged matter. Gaugino condensation then generates the dynamical superpotential
\begin{align}
    W_{\rm dyn} =& 12 \left(\frac{\Lambda^{48}}{\mathcal{I}_4^3}\right)^{1/12}
\end{align}
This is a runaway potential, and so the lower rank orbits corresponding to $\mathcal{I}_4=0$ are not a part of the quantum moduli space. Thus the AMSB theory has residual gauge symmetry $E_6$ with no isolated $U(1)$ factor and so cannot have a massless spin-1 particle in its spectrum.

For completeness I work out the explicit minimum of the AMSB theory. We can take a representative of the $\mathcal{I}_4\neq 0$ orbit as lying purely along the singlet directions, with the D-flat constraint forcing them to be equal. That is, $\phi=(v,0,0,v)$ and so $\mathcal{I}_4 = v^4$. We then have $W_{\rm dyn} = 12 (\Lambda^{48}/v^{12})^{1/12}$ and $K = 2\abs{v}^2$, so that the AMSB potential is
\begin{align}
    V =& \frac{1}{2}\abs{\frac{1}{v}\left(\frac{\Lambda^{48}}{v^{12}}\right)^{1/12}}^2 - 13 m \left(\frac{\Lambda^{48}}{v^{12}}\right)^{1/12} + h.c.
    \\
    =& \frac{1}{2\abs{v}^4}\left(
    \Lambda^8 -52 m \abs{v}^3\Lambda^4 \cos(\arg(v))
    \right)
\end{align}
This is minimized for $v = (\Lambda^4/13m)^{1/3}$.

At the symmetry-breaking scale, the Higgs mechanism gives masses to the gauge bosons in the coset $E_7/E_6$, so one obtains $55$ massive $\mathcal N=1$ vector multiplets with characteristic mass
\begin{align}
m_V \sim g_7 |v| \sim g_7\Lambda \left(\frac{\Lambda}{m}\right)^{1/3}.
\end{align}
There are no remaining massless charged states, and below $m_V$ the gauge group is pure $\mathcal N=1$ $E_6$ SYM, which confines and produces only massive color singlet composites (glueballs/gluinoballs and their excited states).
To estimate the confinement scale, match $W_{\rm dyn}$ to the gaugino-condensation form for $E_6$. Parametrically,
\begin{align}
\Lambda_6 \sim \left(\frac{\Lambda^4}{v}\right)^{1/3}
\sim \Lambda^{8/9}m^{1/9},
\end{align}
up to order-one factors. Therefore the light confined singlets have masses of order $\Lambda_6$:
\begin{align}
m_{\rm glueball}\sim m_{\rm gluinoball}\sim \Lambda_6,.
\end{align}
The modulus along the D-flat direction is light compared to $m_V$, and is stabilized by AMSB to have mass $m_{\rm mod}\sim m$.

\begin{figure}[t]
\centering
\begin{tikzpicture}[x=1cm,y=0.75cm]
  \draw[->] (0,0) -- (0,7.2) node[above] {mass};

  \draw (0,0) -- (-0.15,0) node[left] {$0$};
  \draw (0,1.5) -- (-0.15,1.5) node[left] {$m$};
  \draw (0,4.5) -- (-0.15,4.5) node[left] {$\Lambda$};

  \node[left] at (-0.2,3.2) {$\Lambda_6$};

  \draw[thick] (0.2,1.5) -- (3.8,1.5);
  \node[right,align=left] at (4.0,1.5)
    {D-flat modulus\\$m_{\rm mod}\sim m$};

  \draw[thick] (0.2,3.2) -- (4.6,3.2);
  \node[right,align=left] at (4.8,3.2)
    {confined singlets\\glueball/gluinoball\\scale $\Lambda_6\sim \Lambda^{8/9}m^{1/9}$};

  \draw[thick] (0.2,6.2) -- (5.2,6.2);
  \node[right,align=left] at (5.4,6.2)
    {Higgsed sector\\55 massive vector multiplets\\$m_V\sim g_7\Lambda (\Lambda/m)^{1/3}$};

\end{tikzpicture}
\caption{Parametric mass spectrum for the $E_7$ theory with a rank-4 $\mathcal I_4\neq 0$ AMSB vacuum (residual $E_6$ confinement). It should be understood that there is splitting among the $\mathcal{N}=1$ multiplets.}
\label{fig:E756spectrum}
\end{figure}
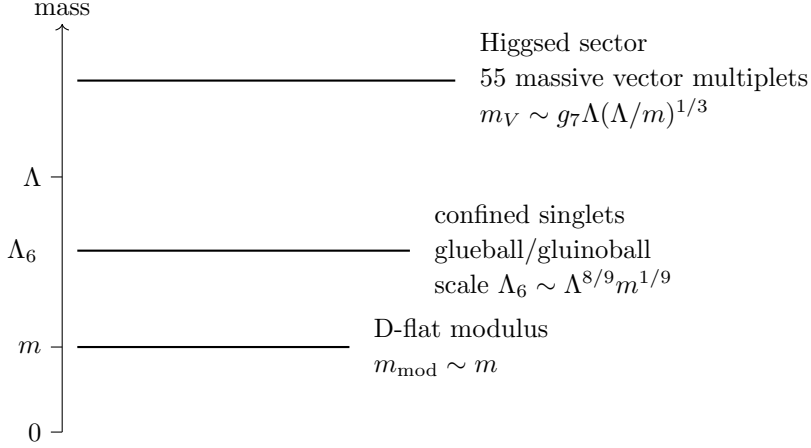

Collecting scales, the hierarchy in the regime $m\ll \Lambda$ is
\begin{align}
m \ll \Lambda_6 \ll \Lambda \lesssim |v| ,
\end{align}
with the Higgsed vectors heavier than $\Lambda$ in typical perturbative regions. Below $m_V$ the $E_6$ vectors confine, and the only light degrees of freedom are the stabilized modulus near $m$ and the singlet glueball/gluinoball composites near $\Lambda_6$, with no massless states. I summarize the parametric form of the spectrum in \cref{fig:E756spectrum}.

\FloatBarrier
\subsection{$\text{Spin}(11)$ with $32_{\rm spin}$}
I follow \cite{cho_exact_1997}. Let $Q$ denote the spinor. At any nonzero modulus the symmetry breaks $\text{Spin}(11)\to SU(5)$ with no isolated $U(1)$. The AMSB theory therefore cannot have any massless spin-1 particles in its spectrum.

For completeness I work out the details of the AMSB vacuum. From \cite{cho_exact_1997} we have
\begin{align}
    W_{\rm dyn} =& 5\left(\frac{\Lambda^{23}}{L^2}\right)^{1/5}
\end{align}
where we have the hadron
\begin{align}
    L =& (Q^T\Gamma_\mu C Q)(Q^T\Gamma^\mu C Q)
\end{align}
$\Gamma$ are the $SO(11)$ Gamma matrices and $C$ is charge conjugation, which we can write as
\begin{equation}
\begin{aligned}
\Gamma_1 &= \sigma_2 \otimes \sigma_3 \otimes \sigma_3 \otimes \sigma_3 \otimes \sigma_3 & \Gamma_2 &= -\sigma_1 \otimes \sigma_3 \otimes \sigma_3 \otimes \sigma_3 \otimes \sigma_3 \\
\Gamma_3 &= 1 \otimes \sigma_2 \otimes \sigma_3 \otimes \sigma_3 \otimes \sigma_3 & \Gamma_4 &= -1 \otimes \sigma_1 \otimes \sigma_3 \otimes \sigma_3 \otimes \sigma_3 \\
\Gamma_5 &= 1 \otimes 1 \otimes \sigma_2 \otimes \sigma_3 \otimes \sigma_3 & \Gamma_6 &= -1 \otimes 1 \otimes \sigma_1 \otimes \sigma_3 \otimes \sigma_3 \\
\Gamma_7 &= 1 \otimes 1 \otimes 1 \otimes \sigma_2 \otimes \sigma_3 & \Gamma_8 &= -1 \otimes 1 \otimes 1 \otimes \sigma_1 \otimes \sigma_3 \\
\Gamma_9 &= 1 \otimes 1 \otimes 1 \otimes 1 \otimes \sigma_2 & \Gamma_{10} &= -1 \otimes 1 \otimes 1 \otimes 1 \otimes \sigma_1
\end{aligned}
\end{equation}
$$\Gamma_{11} = \sigma_3 \otimes \sigma_3 \otimes \sigma_3 \otimes \sigma_3 \otimes \sigma_3 $$
and charge conjugation matrix $C = \sigma_2 \otimes \sigma_1 \otimes \sigma_2 \otimes \sigma_1 \otimes \sigma_2$. 

We can choose a representative of the D-flat $SU(5)$ orbit such that $Q=(v,0,...,0,v)$ with $v\in\mathbb{C}$. Then $L=-4v^4$, $K=2\abs{v}^2$, and $W=5(\Lambda^{23}/16v^8)^{1/5}$ so we can write the AMSB potential as
\begin{align}
    V =&\frac{1}{2}\abs{\frac{8}{v}\left(\frac{\Lambda^{23}}{16 v^{8}}\right)^{1/5}}^2 -13 m \left(\frac{\Lambda^{23}}{16 v^8}\right)^{1/5} + h.c.
    \\
    =& \frac{8 \cdot 2^{2/5} \Lambda^{46/5}}{\vert{}v\vert{}^{26/5}} - \frac{13 \cdot 2^{1/5} m \Lambda^{23/5} \cos\left(\frac{8}{5} \arg(v)\right)}{\vert{}v\vert{}^{8/5}}
\end{align}
This is minimized for
\begin{align}
    v =& 2^{1/3}\left(\frac{\Lambda^{23}}{m^5}\right)^{1/18} e^{\frac{5\pi i k}{4}},\qquad k\in\mathbb{Z}.
\end{align}

\begin{figure}[t]
\centering
\begin{tikzpicture}[x=1cm,y=0.75cm]
  \draw[->] (0,0) -- (0,7.5) node[above] {mass};

  \draw (0,0)   -- (-0.15,0)   node[left] {$0$};
  \draw (0,1.2) -- (-0.15,1.2) node[left] {$m$};
  \draw (0,4.5) -- (-0.15,4.5) node[left] {$\Lambda$};

  \draw (0,2.9) -- (-0.15,2.9) node[left] {$\Lambda_5$};

  \draw[thick] (0.2,1.2) -- (3.4,1.2);
  \node[right,align=left] at (3.6,1.2)
    {one chiral multiplet\ (D-flat modulus)};

  \draw[thick] (0.2,2.9) -- (4.6,2.9);
  \node[right,align=left] at (4.8,2.9)
    {confined singlets:\\glueball/gluinoball composites\\scale $\Lambda_5\sim \Lambda^{23/27}m^{4/27}$};

  \draw[thick] (0.2,6.0) -- (4.9,6.0);
  \node[right,align=left] at (5.1,6.0)
    {Higgsed sector:\\31 massive vector multiplet\\+ heavy $Q$ components\\scale $m_V\sim g_{11} \Lambda (\Lambda/m)^{5/18}$};

\end{tikzpicture}
\caption{Parametric mass spectrum for the $\text{Spin}(11)$ theory with $32_{\rm spin}$ in the AMSB vacuum (residual pure $SU(5)$ confinement). It should be understood that there is splitting among the $\mathcal{N}=1$ multiplets.}
\label{fig:Spin1132spectrum}
\end{figure}
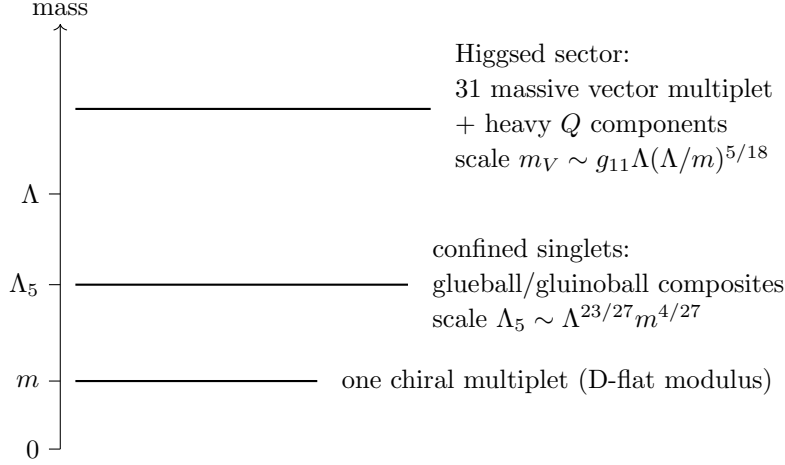

At the Higgs scale $\abs{v}$, the gauge symmetry breaking $\text{Spin}(11)\to SU(5)$ produces massive vectors in the coset $\text{Spin}(11)/SU(5)$, so there are 31 massive $\mathcal{N}=1$ vector multiplets with characteristic mass
\begin{align}
    m_V\sim& g_{11}\abs{v} \sim g_{11}\Lambda\left(\frac{\Lambda}{m}\right)^{5/18}
\end{align}
In addition, after Higgsing one physical complex modulus remains corresponding to the D-flat direction, whose mass is given by AMSB as
\begin{align}
    m_{\rm mod}\sim& m.
\end{align}
Below $m_V$, the remaining gauge theory is pure $\mathcal{N}=1$ $SU(5)$ SYM and confines. Matching $W_{\rm dyn}=5\Lambda_5^3$ gives
\begin{align}
    \Lambda_5\sim& \Lambda^{23/27}m^{4/27}. 
\end{align}
The confined singlet spectrum (glueballs/gluinoballs and excited states) then has characteristic masses
\begin{align}
    m_{\rm glueball} \sim& m_{\rm gluinoball} \sim \Lambda_5.
\end{align}
There are no massless vectors because $SU(5)$ confines and no isolated $U(1)$ remains. I summarize the parametric form of the spectrum in \cref{fig:Spin1132spectrum}.

\FloatBarrier
\subsection{$\text{Spin}(12)$ with $32_{\rm spin}$}
Note that the $32_{\rm spin}$ featured in \cite{cacciapaglia_perusing_2026} is the one corresponding to the odd exterior power. Following \cite{maru_confining_1998} let $Q$ denote the spinor\footnote{Note that \cite{maru_confining_1998} explicitly uses the other $32$ dimensional spinor irrep, but the same conclusions can be applied since the theories are related by an outer automorphism and as such are locally equivalent.}. The D-flat constraint breaks $\text{Spin}(12)\to SU(6)$ for any nonzero modulus, with no isolated $U(1)$. The nonabelian factor confines. Thus the AMSB theory cannot have a massless spin-1 particle in its spectrum.

For completeness I work out the details of the AMSB vacuum. Following \cite{maru_confining_1998} we have 
\begin{align}
    W_{\rm dyn} =& 6 \left(\frac{\Lambda^{26}}{L^2}\right)^{1/6}
\end{align}
with 
\begin{equation}
    L = \frac{1}{4}(Q^T \Gamma^{[\mu}\Gamma^{\nu]}CQ)(Q^T \Gamma_{[\mu}\Gamma_{\nu]}CQ)
\end{equation}
where
\begin{equation}
\begin{aligned}
\Gamma_1 &= \sigma_2 \otimes \sigma_3 \otimes \sigma_3 \otimes \sigma_3 \otimes \sigma_3 \otimes \sigma_3 & \Gamma_2 &= -\sigma_1 \otimes \sigma_3 \otimes \sigma_3 \otimes \sigma_3 \otimes \sigma_3 \otimes \sigma_3 \\
\Gamma_3 &= 1 \otimes \sigma_2 \otimes \sigma_3 \otimes \sigma_3 \otimes \sigma_3 \otimes \sigma_3 & \Gamma_4 &= -1 \otimes \sigma_1 \otimes \sigma_3 \otimes \sigma_3 \otimes \sigma_3 \otimes \sigma_3 \\
\Gamma_5 &= 1 \otimes 1 \otimes \sigma_2 \otimes \sigma_3 \otimes \sigma_3 \otimes \sigma_3 & \Gamma_6 &= -1 \otimes 1 \otimes \sigma_1 \otimes \sigma_3 \otimes \sigma_3 \otimes \sigma_3 \\
\Gamma_7 &= 1 \otimes 1 \otimes 1 \otimes \sigma_2 \otimes \sigma_3 \otimes \sigma_3 & \Gamma_8 &= -1 \otimes 1 \otimes 1 \otimes \sigma_1 \otimes \sigma_3 \otimes \sigma_3 \\
\Gamma_9 &= 1 \otimes 1 \otimes 1 \otimes 1 \otimes \sigma_2 \otimes \sigma_3 & \Gamma_{10} &= -1 \otimes 1 \otimes 1 \otimes 1 \otimes \sigma_1 \otimes \sigma_3 \\
\Gamma_{11} &= 1 \otimes 1 \otimes 1 \otimes 1 \otimes 1 \otimes \sigma_2 & \Gamma_{12} &= -1 \otimes 1 \otimes 1 \otimes 1 \otimes 1 \otimes \sigma_1
\end{aligned}
\end{equation}
$$\Gamma_{13} = \sigma_3 \otimes \sigma_3 \otimes \sigma_3 \otimes \sigma_3 \otimes \sigma_3 \otimes \sigma_3 $$
and $C$ is charge conjugation.

\begin{figure}[t]
\centering
\begin{tikzpicture}[x=1cm,y=0.75cm]
  \draw[->] (0,0) -- (0,7.5) node[above] {mass};

  \draw (0,0)   -- (-0.15,0)   node[left] {$0$};
  \draw (0,1.2) -- (-0.15,1.2) node[left] {$m$};
  \draw (0,4.5) -- (-0.15,4.5) node[left] {$\Lambda$};

  \draw (0,2.9) -- (-0.15,2.9) node[left] {$\Lambda_6$};

  \draw[thick] (0.2,1.2) -- (3.4,1.2);
  \node[right,align=left] at (3.6,1.2)
    {one chiral multiplet\ (D-flat modulus)};

  \draw[thick] (0.2,2.9) -- (4.6,2.9);
  \node[right,align=left] at (4.8,2.9)
    {confined singlets:\\glueball/gluinoball composites\\scale $\Lambda_6\sim \Lambda^{13/15}m^{2/15}$};

  \draw[thick] (0.2,6.0) -- (4.9,6.0);
  \node[right,align=left] at (5.1,6.0)
    {Higgsed sector:\\31 massive vector multiplets\\+ heavy $Q$ components\\scale $m_V\sim g_{12} \Lambda (\Lambda/m)^{3/10}$};

\end{tikzpicture}
\caption{Parametric mass spectrum for the $\text{Spin}(12)$ theory with $32_{\rm spin}$ in the AMSB vacuum (residual pure $SU(6)$ confinement). It should be understood that there is splitting among the $\mathcal{N}=1$ multiplets.}
\label{fig:Spin1232spectrum}
\end{figure}
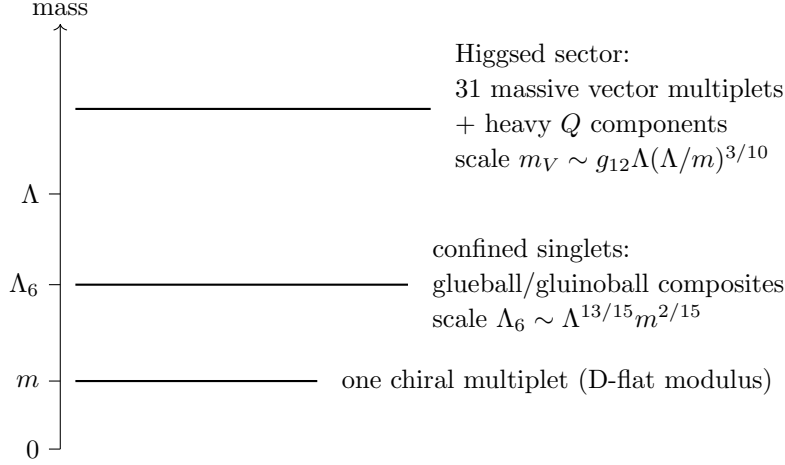

We can choose a represnetative of the $SU(6)$ orbit as $Q=(0,v,0,...,0,v,0)$. Then $L=12v^4$ and $W=6(\Lambda^{26}/144v^8)^{1/6}$ while the K\"ahler potential is $2\abs{v}^2$. So the AMSB potential is
\begin{align}
    V =& \frac{1}{2}\abs{\frac{8}{v}\left(\frac{\Lambda^{26}}{144 v^8}\right)^{1/6}}^2 -12m\left(\frac{\Lambda^{26}}{144 v^8}\right)^{1/6} + h.c.
    \\
    =& 8\left(\frac{4\Lambda^{26}}{9\abs{v}^{14}}\right)^{1/3} - 4 m \left(\frac{18 \Lambda^{13}}{\abs{v}^4}\right)^{1/3}\cos(\frac{4}{3}\arg(v))
\end{align}
This is minimized for
\begin{align}
    v =& \left(\frac{686\Lambda^{13}}{81m^3}\right)^{1/10}
    e^{\frac{3\pi i}{2}k},\qquad k\in\mathbb{Z}
\end{align}
The discrete family of minima corresponds to the branch structure induced by the $1/6$ power in $W_{\rm dyn}$ (ultimately tied to the discrete chiral/center structure of the confining $SU(6)$ sector).
At the Higgs scale $|v|$, the gauge symmetry breaking $\text{Spin}(12)\to SU(6)$ produces massive vectors in the coset $\text{Spin}(12)/SU(6)$, 
so there are $31$ massive $\mathcal N=1$ vector multiplets with characteristic mass
\begin{align}
m_V\sim g_{12}|v| \sim g_{12}\Lambda \left(\frac{\Lambda}{m}\right)^{3/10}.
\end{align}
In addition, after Higgsing one physical complex modulus remains (the D-flat direction represented by $v$), whose mass is parametrically
\begin{align}
m_{\rm mod}\sim m.
\end{align}
Below $m_V$, the remaining gauge theory is pure $\mathcal N=1$ $SU(6)$ SYM and confines. Matching $W_{\rm dyn}=6\Lambda_6^3$ to the given expression yields
\begin{align}
\Lambda_6 \sim \Lambda^{13/15}m^{2/15}.
\end{align}
Thus the confined singlet spectrum (glueballs/gluinoballs and excited states) has characteristic masses
\begin{align}
m_{\rm glueball}\sim m_{\rm gluinoball}\sim \Lambda_6.
\end{align}
There are no massless vectors because $SU(6)$ confines and no isolated $U(1)$ remains. I summarize the parametric form of the spectrum in \cref{fig:Spin1232spectrum}.

\FloatBarrier
\subsection{$\text{Spin}(13)$ with $64_{\rm spin}$}
\begin{figure}
\centering
\begin{tikzpicture}[
    scale=2.5,
  font=\small,
  special/.style={line width=1.1pt, dashed},
  lab/.style={
    draw, rounded corners, align=left,
    inner sep=2.3pt, font=\scriptsize,
    text width=3.4cm, fill=white, opacity=0.95
  }
]

  \draw[->, line width=0.9pt] (0,0) -- (3.2,0) node[right] {$\alpha$};
  \draw[->, line width=0.9pt] (0,0) -- (0,3.2) node[above] {$\beta$};

  \foreach \x in {0.5,1,1.5,2,2.5,3} \draw[gray] (\x,0.08) -- (\x,-0.08);
  \foreach \y in {0.5,1,1.5,2,2.5,3} \draw[gray] (0.08,\y) -- (-0.08,\y);

  \draw[special, blue!75!black] (0.12,0.12) -- (2.95,2.95);   

  \node[lab, fill=orange!12] at (-0.15,2.55) [anchor=west] {
    $\alpha=0,\ \beta\neq 0$\\
    $SO(13)\rightarrow SU(6)$ with $\ydiagram{1,1,1}$
  };
  \node[lab, fill=orange!12] at (2.1,-0.15) [anchor=south] {
    $\alpha\neq 0,\ \beta=0$\\
    $SO(13)\rightarrow SU(6)$ with $\ydiagram{1,1,1}$
  };

  \node[lab, fill=blue!10] at (1.65,1.95) [anchor=south west] {
    $\alpha=\beta\neq 0$\\
    $SO(13)\rightarrow SU(3)\times G_2$\\
    one field: $(\mathbf{1},\mathbf{7})=\alpha-\beta$
  };

  \node[lab, fill=gray!12] at (2.15,0.80) [anchor=south west] {
    generic $\alpha,\beta\neq0,\ \alpha\neq\beta$\\
    $SO(13)\rightarrow SU(3)\times SU(3)$\\
    no matter charged under the unbroken group
  };

\end{tikzpicture}
\caption{Symmetry breaking pattern of $\mathcal{N}=1$ $SO(13)$ with a $64_{\rm spinor}$ depending on the moduli $\alpha$ and $\beta$. Each possibility eventually breaks down to pure $SU(3)\times SU(3)$ in the deep IR. The scalar potential is singular along the $\alpha=0,\beta=0,\alpha-\beta=0$ lines. AMSB lifts the runaway and leads to a stable minimum in the generic $\alpha\neq \beta$ region.}
\label{fig:SO(13)alphabeta}
\end{figure}
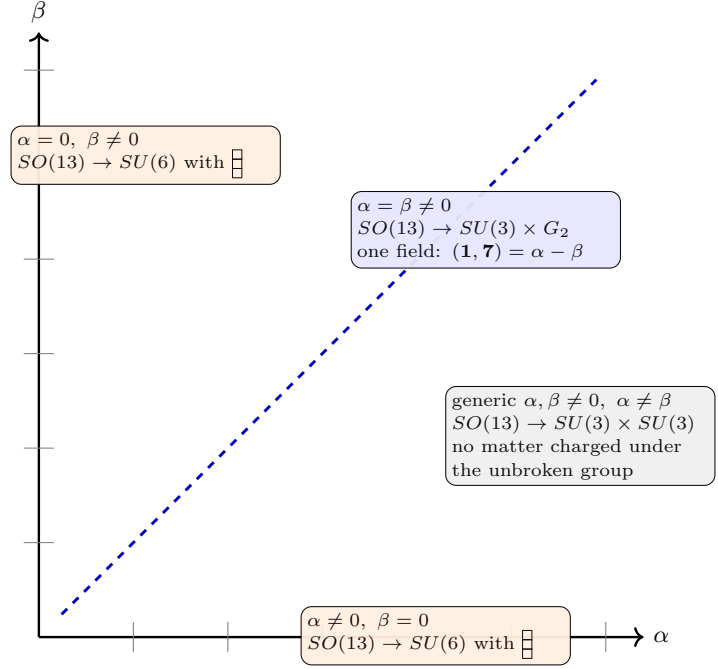

From \cite{dotti_free_1998} we have
\begin{align}
    W_{\rm dyn} =& \frac{\Lambda^{25/3}}{X^{1/3}Y^{1/2}}\left(\omega^i-\omega^j\frac{X^{2/3}}{(X^2-4Y)^{1/3}}\right) \label{eq:WdynSpin13with64}
\end{align}
where $\omega = e^{2\pi i /3}$, ${i,j}$ denote branches of the cube root, and with $S$ the spinor we have $X=S^4$ and $Y=S^8$. Just as in \cite{dotti_free_1998} we take $S=\alpha S_1+\beta S_2$ to parameterize the inequivalent flat directions, and as such $X=\alpha^4+\beta^4$ and $Y=\alpha^4\beta^4$. The symmetry breaking pattern depends on the relative values of $\alpha$ and $\beta$ and is depicted in \cref{fig:SO(13)alphabeta}, with all options generating dynamical superpotentials covered by \cref{eq:WdynSpin13with64}. On the branches where $SO(13)\to SU(6)$ there is a massless three-index antisymmetric tensor, and so the dynamics proceeds as in \cref{sec:SU6withA3} toward a pure $SU(3)\times SU(3)$. Generic $\alpha\neq \beta$ break $SO(13)\to SU(3)\times SU(3)$ with no massless matter directly. Finally, the $\alpha=\beta\neq 0$ branch breaks $SO(13)\to SU(3)\times G_2$ with a $(1,7)$, and $G_2$ with a $7$ breaks down to $SU(3)$. So in all cases the low energy theory is pure $SU(3)\times SU(3)$, with no isolated $U(1)$ and so no possibility of a massless spin-1 particle in the AMSB spectrum. Of course, the superpotential above is singular on those special loci and so those points are removed from the full quantum theory.

For completeness I calculate the AMSB minimum explicitly. Since $S_1\perp S_2$, $K=\abs{\alpha}^2+\abs{\beta}^2$ and so the AMSB scalar potential is
\begin{align}
    V =& \abs{\dpd{W}{\alpha}}^2+\abs{\dpd{W}{\beta}}^2 + m (\alpha\dpd{W}{\alpha}+\beta\dpd{W}{\beta}-3W) + h.c.
\end{align}
Note that $W(t\alpha,t\beta) = t^{-16/3} W(\alpha,\beta)$, so from Euler's theorem we have
\begin{align}
    \alpha \dpd{W}{\alpha} + \beta\dpd{W}{\beta} =& -\frac{16}{3}W,
\end{align}
and so we can re-write the AMSB terms as
\begin{align}
    V_{\rm AMSB} =& \frac{25}{16}m(\alpha \dpd{W}{\alpha}+\beta\dpd{W}{\beta} + h.c.).
\end{align}
I then complete the square with the F-terms to get
\begin{align}
    V =& \abs{\dpd{W}{\alpha}+\frac{25}{16}m\alpha^*}^2 + \abs{\dpd{W}{\beta}+\frac{25}{16}m\beta^*}^2 - \left(\frac{25}{16}\right)^2m^2\left(\abs{\alpha}^2+\abs{\beta}^2\right).
\end{align}
The minimum will occur when the first two terms vanish. Taking
\begin{align}
W=&\Lambda^{25/3}\,\alpha^{-16/3}\,G_{ij}(r),&
\text{ where }\beta = r\alpha\text{ and }
G_{ij}(r)=&
\frac{ \omega^i-\omega^j\left(\frac{1+r^4}{1-r^4}\right)^{2/3}}
{ r^2(1+r^4)^{1/3}},
\end{align}
the solution can be written as
\begin{align}
|\alpha_{ij}|^{22/3}=&\dfrac{|\Lambda|^{25/3}|G_{ij}'(r_{ij})|}{\frac{25}{16}|m||r_{ij}|},
\end{align}
where $r_{ij}$ is the solution to
\begin{align}
    r_{ij}^*\Big[-\frac{16}{3}G_{ij}(r)-r_{ij}G_{ij}'(r_{ij})\Big]=&G_{ij}'(r_{ij}).
\end{align}
The phases of $\alpha_{ij}$ and $\beta_{ij}=r_{ij}\alpha_{ij}$ (no summation) are then determined by plugging back in to the first two terms and solving. In lieu of an analytic solution, I numerically solve for the $r_{ij}$. The results show $\abs{r_{ij}}\sim \mathcal{O}(1)$ and $\abs{G'_{ij}(r_{ij})}\sim\mathcal{O}(1)$. So for our purposes here I just report the scaling behavior and leave more detailed analysis to future work. We thus have that $\abs{\alpha_{ij}}\sim\abs{\beta_{ij}}\sim\mathcal{O}(1) \times\Lambda\left(\frac{16 \Lambda}{25 m}\right)^{3/22}$ and $V_{\rm min}\sim - \frac{625}{128}m^2\Lambda^2 \left(\frac{16\Lambda}{25 m}\right)^{3/11}$

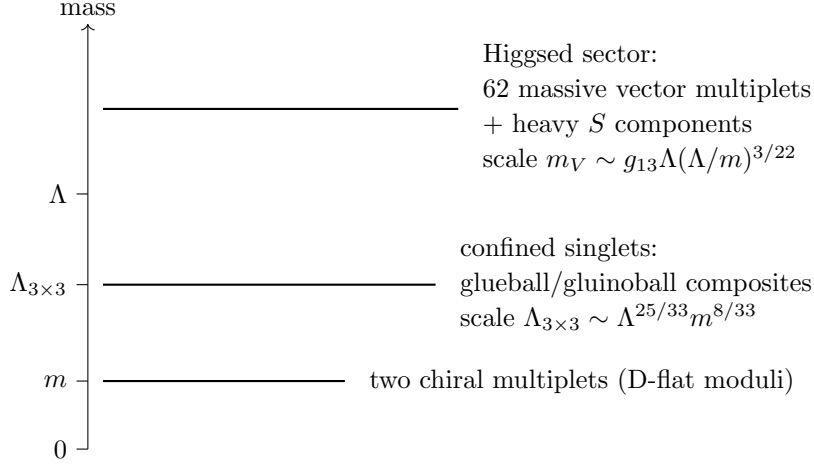
\begin{figure}[t]
\centering
\begin{tikzpicture}[x=1cm,y=0.75cm]
  \draw[->] (0,0) -- (0,7.5) node[above] {mass};

  \draw (0,0)   -- (-0.15,0)   node[left] {$0$};
  \draw (0,1.2) -- (-0.15,1.2) node[left] {$m$};
  \draw (0,4.5) -- (-0.15,4.5) node[left] {$\Lambda$};

  \draw (0,2.9) -- (-0.15,2.9) node[left] {$\Lambda_{3\times 3}$};

  \draw[thick] (0.2,1.2) -- (3.4,1.2);
  \node[right,align=left] at (3.6,1.2)
    {two chiral multiplets\ (D-flat moduli)};

  \draw[thick] (0.2,2.9) -- (4.6,2.9);
  \node[right,align=left] at (4.8,2.9)
    {confined singlets:\\glueball/gluinoball composites\\scale $\Lambda_{3\times 3}\sim \Lambda^{25/33}m^{8/33}$};

  \draw[thick] (0.2,6.0) -- (4.9,6.0);
  \node[right,align=left] at (5.1,6.0)
    {Higgsed sector:\\62 massive vector multiplets\\+ heavy $S$ components\\scale $m_V\sim g_{13} \Lambda (\Lambda/m)^{3/22}$};

\end{tikzpicture}
\caption{Parametric mass spectrum for the $\text{Spin}(13)$ theory with $64_{\rm spin}$ in the AMSB vacuum (residual pure $SU(3)\times SU(3)$ confinement). It should be understood that there is splitting among the $\mathcal{N}=1$ multiplets.}
\label{fig:Spin1364spectrum}
\end{figure}

The Higgs scale is therefore $\sim \Lambda (\Lambda/m)^{3/22}$, where the gauge symmetry breaks $\text{Spin(13)}\to SU(3)\times SU(3)$ producing massive vectors in the coset $\text{Spin}(13)/SU(3)\times SU(3)$, so there are 62 massive $\mathcal{N}=1$ vector multiplets with characteristic mass
\begin{align}
    m_V \sim&\, g_{13} \Lambda\left(\frac{\Lambda}{m}\right)^{3/22}.
\end{align}
In addition, after Higgsing there are two chiral multiplets corresponding to the D-flat directions whose masses come solely from AMSB and scales as
\begin{align}
    m_{\rm mod}\sim& m.
\end{align}
Below $m_V$, the remaining gauge theory is pure $\mathcal{N}=1$ $SU(3)\times SU(3)$ SYM and confines at the scale
\begin{align}
    \Lambda_{3\times 3} \sim&\, m^{8/33}\Lambda^{25/33}.
\end{align}
It follows that the confined singlet spectrum (glueballs/gluinoballs and excited states) has a characteristic mass
\begin{align}
    m_{\rm glueball}\sim& m_{\rm gluinoball} \sim \Lambda_{3\times 3}.
\end{align}
There are no massless vectors because $SU(3)\times SU(3)$ confines and no isolated $U(1)$ remains. I summarize the parametric form of the spectrum in \cref{fig:Spin1364spectrum}.

\FloatBarrier
\subsection{$Sp(6)$ with $6_F+14'_{A3}$}\label{sec:Sp6FA3}
Let $Q$ be the fundamental and $B$ the $\ydiagram{1,1,1}$. We use the symplectic invariant $J^{\alpha\beta} = (1_{3\times 3}\otimes i\sigma_2)^{\alpha\beta}$ to raise indices (i.e. $Q^\tau = J^{\tau\delta}Q_\delta$). The gauge invariant moduli are $B^4$ and $B^2Q^2$ where
\begin{align}
    \bar{B}^{\alpha\beta\gamma} =& B_{\mu\nu\tau}\epsilon^{\alpha\beta\gamma\mu\nu\tau}
    \\
    B^4 \equiv& \bar{B}^{\alpha\beta\gamma}\bar{B}^{\mu\nu\tau}B_{\alpha\beta\tau}B_{\mu\nu\gamma}
    \\
    B^2Q^2 \equiv& \bar{B}^{\alpha\beta\gamma}B_{\alpha\beta\tau}Q^\tau Q_\gamma 
\end{align}
The theory also has a non-anomalous flavor symmetry $U(1)_F$, under which
\begin{align}
    q_F(Q)=5,
    \qquad
    q_F(B)=-1.
\end{align}
The corresponding charges of the gauge-invariant coordinates are
\begin{align}
    q_F(B^4)=-4,
    \qquad
    q_F(B^2Q^2)=8.
\end{align}
Consequently, the dynamically generated superpotential below is invariant,
since
\begin{align}
    q_F\!\left(B^4\sqrt{B^2Q^2}\right)
    =-4+\frac12(8)=0.
\end{align}
The AMSB deformation also preserves this non-$R$ flavor symmetry.

The D-flat condition can be re-written as the vanishing of the symmetric part of a combination of bilinears like so:
\begin{align}
    Q^{\dagger (i} Q^{j)} \;+\; \frac12\, B^{\dagger (i|kl|}\, B^{j)}{}_{kl}\;=\;0
\end{align}
A generic D-flat point on the moduli space breaks $Sp(6)\to SU(2)$ with no isolated $U(1)$, as in \cite{dotti_free_1998}. There are enhanced symmetry directions where one of the invariants vanish, but those also do not lead to isolated $U(1)$'s and moreover are absent from the quantum theory as we will see below. Thus the AMSB theory cannot have a massless spin-1 particle in its spectrum.

Because there are only two independent gauge invariants, the complex dimension of the moduli space should be two. We should be able to parametrize a generic point in moduli space, after fully gauge fixing, in terms of two complex numbers. We will take the standard normalization of the generators $\text{Tr}(T^aT^b)=\delta^{ab}/2$.

For completeness I explicitly minimize the AMSB theory. Choose parameters $x,y,z\in\mathbb{C}$ all nonzero. I find one representative D-flat VEV to be
\begin{align}
Q_\alpha =& (\sqrt{2}x,0,0,0,0,0) \\
B_{123} =& B_{145} = B_{356} = y,\qquad B_{235} = x,\qquad B_{246} = z,
\end{align}
with all other independent components set to zero. D-flatness constrains $\abs{z}^2 = \abs{x}^2+\abs{y}^2$, but I do not implement this directly in order to preserve the manifest holomorphicity of the superpotential. In this parametrization we have
\begin{align}
    K =& \abs{z}^2 + 3\abs{x}^2 + 3\abs{y}^2,
    \\
    B^2Q^2 =& 48 x^3 z,
    \\
    B^4 =& 3456 y^3 z.
\end{align}
Although the flavor transformation does not act diagonally on the particular
gauge-fixed field representative above, its action on the moduli-space
coordinates is unambiguous. In a gauge in which $z$ is chosen real, it can
be represented as
\begin{align}
    x &\longrightarrow e^{8i\alpha/3}x,
    &
    y &\longrightarrow e^{-4i\alpha/3}y,
    &
    z &\longrightarrow z,
\end{align}
which indeed gives
\begin{align}
    B^2Q^2 &\longrightarrow e^{8i\alpha}B^2Q^2,
    &
    B^4 &\longrightarrow e^{-4i\alpha}B^4.
\end{align}
From \cite{dotti_free_1998} we have a dynamically generated superpotential
\begin{align}
    W_{\rm dyn} =& \pm \frac{2i\Lambda^9}{B^4\sqrt{B^2Q^2}},
\end{align}
where $\pm$ denotes two branches. In the above parametrization this is
\begin{align}
    W_{\rm dyn} =& \pm\frac{i\Lambda^9}{2^8 3^3 y^3 z \sqrt{3x^3 z}}
\end{align}
Because I have left $z$ free, we must include in the scalar potential its D-term. The full scalar potential including that D-term and AMSB is then
\begin{align}
    V =& \abs{\dpd{W}{z}}^2 + \frac{1}{3}\abs{\dpd{W}{x}}^2 + \frac{1}{3}\abs{\dpd{W}{y}}^2
    + 
   V_D
    + m (x\dpd{W}{x}+y\dpd{W}{y}+z\dpd{W}{z}-3W) + h.c.
\end{align}
I have already been approximating that the minimum should be D-flat, so it is okay to impose after-the-fact\footnote{In fact, this is what is done in virtually every AMSB paper. Deviations from the D-flat loci must vanish parametrically in $m\to 0$. Together with the fact that $V_D$ is positive-definite, taking the AMSB minimum to be D-flat should generically be a good approximation. The question of whether the AMSB minimum can ever deviate largely from the D-flat loci is an interesting one, but not likely to be relevant in most cases and beyond the scope of this work.}. So I set $V_D=0$, compute the other terms, and then set $z=\sqrt{\abs{x}^2+\abs{y}^2}$. Doing so gives
\begin{align}
    V =&  
    \frac{\Lambda^{18}}{2^{18}3^6 \abs{x^5y^8z^7}}\left(3 | x| ^2 | y| ^2 | z| ^2+4 | x| ^2 | z| ^4+| y| ^2 | z| ^4\right)
    \mp m \frac{i\Lambda^9 x^3}{768\sqrt{3} y^3 (x^3 z)^{3/2}} + h.c.
    \\
    =& 
    \frac{\Lambda ^{18} \left(4 r_x^4+8 r_x^2 r_y^2+r_y^4\right)}{2^{18}3^6 r_x^5 r_y^8 \left(r_x^2+r_y^2\right)^{5/2}}\mp m \frac{\Lambda ^9 r_x^3 \sin\left(\frac{3}{2}\phi_x+3\phi_y\right)}{384 \sqrt{3} r_y^3 \left(r_x^3  \sqrt{r_x^2+r_y^2}\right)^{3/2}} 
\end{align}
where I've taken $x=r_xe^{i\phi_x}$ and $y= r_y e^{i\phi_y}$. The potential depends on the phases only through the flavor-invariant
combination
\begin{align}
    \chi=\frac32\phi_x+3\phi_y.
\end{align}
For each branch, $\chi$ is fixed so as to make the AMSB contribution as
negative as possible. The orthogonal phase combination is not fixed.
Indeed, under $U(1)_F$,
\begin{align}
    \delta\phi_x=\frac83\alpha,
    \qquad
    \delta\phi_y=-\frac43\alpha,
\end{align}
so that $\delta\chi=0$, whereas, for example,
\begin{align}
    \eta\equiv\phi_x-\phi_y
    \qquad\Longrightarrow\qquad
    \delta\eta=4\alpha.
\end{align}
The minimum found below is therefore a continuous $U(1)_F$ orbit of
degenerate vacua rather than an isolated minimum. Then I find the global minimum at
\begin{align}
    r_x =& \frac{\sqrt[8]{7} \Lambda ^{9/8}}{2\ 2^{5/16} 3^{25/32} \sqrt[8]{m}}
    &
    r_y =& \frac{\sqrt[8]{7} \Lambda ^{9/8}}{2^{13/16} 3^{25/32} \sqrt[8]{m}}
\end{align}
with $V_{\rm min} = -\frac{18\ 2^{3/8} 3^{7/16} \Lambda ^{9/4} m^{7/4}}{7\ 7^{3/4}}\approx -1.25 m^{7/4}\Lambda^{9/4}$. Since both $B^4$ and $B^2Q^2$ are nonzero at this minimum and both carry
nonzero $U(1)_F$ charge, the vacuum spontaneously breaks $U(1)_F$ to a
discrete subgroup. It therefore contains one exactly massless
Nambu--Goldstone boson. In the coordinates above, the Nambu--Goldstone mode
is the phase fluctuation along the $\eta$ direction. Its decay constant is
parametrically of order the Higgs scale,
\begin{align}
    f_{\rm NGB}
    \sim \Lambda\left(\frac{\Lambda}{m}\right)^{1/8}.
\end{align}
At this scale the gauge symmetry breaks $Sp(6)\to SU(2)$. The vectors in the
coset $Sp(6)/SU(2)$ therefore form 18 massive $\mathcal N=1$ vector
multiplets, up to the splittings induced by AMSB, with characteristic mass
\begin{align}
    m_V\sim g_6\Lambda\left(\frac{\Lambda}{m}\right)^{1/8}.
\end{align}

Before the AMSB deformation, the D-flat moduli space has two complex
dimensions. The deformation stabilizes the two radial directions and the
flavor-invariant angular direction, with characteristic masses of order
$m$. The remaining angular direction is the Nambu--Goldstone boson of the
spontaneously broken $U(1)_F$ and is exactly massless. Since supersymmetry
is softly broken, its fermionic partner is not protected by Goldstone's
theorem and is generically massive, with a characteristic mass of order
$m$.

Below $m_V$, the remaining gauge theory is pure $\mathcal N=1$
$SU(2)$ SYM, which confines at the scale
\begin{align}
    \Lambda_2\sim m^{1/4}\Lambda^{3/4}.
\end{align}
The confined gauge-sector singlets, including glueball and gluinoball
states, have characteristic masses of order $\Lambda_2$. Confinement
does not gap the flavor Nambu--Goldstone boson, because $U(1)_F$ is an
exact non-anomalous symmetry and is broken by the VEVs of the charged
gauge-invariant operators $B^4$ and $B^2Q^2$.

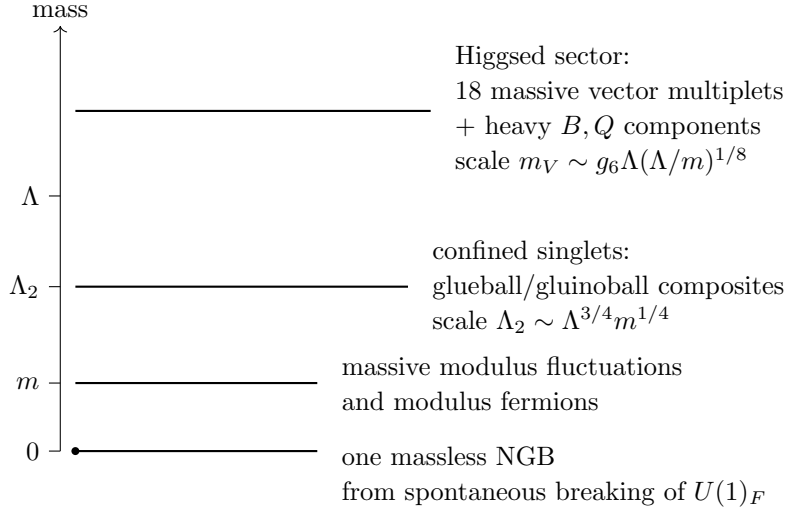
\begin{figure}[t]
\centering
\begin{tikzpicture}[x=1cm,y=0.75cm]
  \draw[->] (0,0) -- (0,7.5) node[above] {mass};

  \draw (0,0)   -- (-0.15,0)   node[left] {$0$};
  \draw (0,1.2) -- (-0.15,1.2) node[left] {$m$};
  \draw (0,4.5) -- (-0.15,4.5) node[left] {$\Lambda$};
  \draw (0,2.9) -- (-0.15,2.9) node[left] {$\Lambda_{2}$};

  \fill (0.2,0) circle (1.5pt);
  \draw[thick] (0.2,0) -- (3.4,0);
  \node[right,align=left] at (3.6,0)
    {\\ \\ one massless NGB\\from spontaneous breaking of $U(1)_F$};

  \draw[thick] (0.2,1.2) -- (3.4,1.2);
  \node[right,align=left] at (3.6,1.2)
    {massive modulus fluctuations\\and modulus fermions};

  \draw[thick] (0.2,2.9) -- (4.6,2.9);
  \node[right,align=left] at (4.8,2.9)
    {confined singlets:\\glueball/gluinoball composites\\
     scale $\Lambda_{2}\sim \Lambda^{3/4}m^{1/4}$};

  \draw[thick] (0.2,6.0) -- (4.9,6.0);
  \node[right,align=left] at (5.1,6.0)
    {Higgsed sector:\\18 massive vector multiplets\\
     $+$ heavy $B,Q$ components\\
     scale $m_V\sim g_{6}\Lambda(\Lambda/m)^{1/8}$};

\end{tikzpicture}
\caption{Parametric mass spectrum for the $Sp(6)$ theory with
$6_F+14'_{A3}$ in the AMSB vacuum, with residual pure $SU(2)$
confinement. The exact non-anomalous flavor symmetry $U(1)_F$ is
spontaneously broken, producing one massless Nambu--Goldstone boson.
The remaining modulus fluctuations and their fermionic partners are
generically massive at the AMSB scale. It should be understood that
AMSB induces splittings among states inherited from the
$\mathcal N=1$ multiplets.}
\label{fig:Sp(6)6+14spectrum}
\end{figure}

Thus the spectrum contains one exactly massless scalar Nambu--Goldstone
boson, but no massless spin-1 state: the residual $SU(2)$ confines and no
isolated gauge $U(1)$ remains. The parametric spectrum is summarized in
\cref{fig:Sp(6)6+14spectrum}.

\FloatBarrier
\section{In the CW?}\label{sec:InCW}

Of the 19 pseudoreal theories considered in \cite{cacciapaglia_perusing_2026}, 13 were found to either ``likely'' be in the CW or it was indeterminate. Of those, four have asymptotically free $\mathcal{N}=1$ SUSY versions. I find that all four of these theories do not admit any dynamically generated superpotentials and are consistent with flowing to an interacting SCFT in the IR (though I do not attempt to prove this). 

This at least does not contradict the claim that the non-SUSY theories may be in the conformal window. However, the AMSB deformation has been shown to in some cases deflect the RG flow away from a superconformal fixed point \cite{kondo_broken_2025}. It is not beyond the realm of possibility that a similar phenomenon could occur in these cases. However the lack of a weakly-coupled description makes such an analysis difficult, if not impossible. I leave any such attempts to future work.

Each of the theories below does not appear to be treated explicitly in the existing SUSY literature to the best of my knowledge. So here I initiate their study in earnest. Before determining if these theories could consistently flow to SCFTs in the IR, I first check that they do not obviously do something else. To do so, I compute the index $\iota\equiv \sum_i\mu_i - \mu_G$ for each theory, where $\mu_i$ is the Dynkin index of the matter representation $R_i$ (normalized so that $\mu_{\ydiagram{1}} = 1$) and $\mu_G$ is that of the adjoint. This index was used in \cite{grinsteinSystematicStudyTheories1998a,grinsteinSystematicStudyTheories1998b,dotti_free_1998,csaki_systematic_1997}, among others, to classify different types of $\mathcal{N}=1$ theories.

The values of the index for the four theories considered below are given in \cref{tab:indices}. Importantly, every theory considered here is of the sort $\iota >2$. This rules out s-confinement, a quantum modified moduli space, and a dynamically generated superpotential consistent with the weak coupling $\Lambda\to 0$ limit. This is consistent with the fact that none of these theories have shown up in the aforementioned systematic studies of such classes. One can also note, due to a result from \cite{andreevOrbitsGreatestDimension1967}, that in each of these theories a generic point in the moduli space breaks the gauge group completely. 

\begin{table}[]
    \centering
\begin{tabular}{c|c}
\text{Theory} & $\iota\equiv \sum_i \mu_i-\mu_G$\\ \hline
$Sp(10)$:\ $10_F+110_{A_3}$ & 16\\
$Sp(4)$:\ $16_{R_{11}}$ & 6\\
$Sp(8)$:\ $48_{A_3}$ & 4\\
$Sp(12)$:\ $203_{A_3}$ & 30
\end{tabular}
    \caption{Indices of the four theories considered to potentially be in the conformal window. Since for all theories $\iota>2$, they do not fall into the classes of $\mathcal{N}=1$ theories that s-confine, have quantum-modified moduli spaces, or develop dynamical superpotentials.}
    \label{tab:indices}
\end{table}

In the remainder of this section, I consider each of the four theories and gather necessary but not sufficient evidence that they flow to SCFTs in the IR. In particular, I identify a consistent non-anomalous $U(1)_R$ in each case that satisfies the a-theorem, is consistent with conformal collider bounds, and makes the NSVZ beta function vanish.

\FloatBarrier
\subsection{SCFT consistency checks}
If a theory flows to an interacting $\mathcal{N}=1$ SCFT, it must possess an anomaly-free superconformal $U(1)_R$ symmetry. For any candidate trial $U(1)_R$, the central charges are fixed by the ’t Hooft anomalies of that $U(1)_R$. Concretely, writing $\mathrm{Tr}$ as the trace over all Weyl fermions in the theory, the conformal anomaly coefficients are
\begin{equation}
a=\frac{3}{32}\Big(3\,\mathrm{Tr}R^3-\mathrm{Tr}R\Big),\qquad
c=\frac{1}{32}\Big(9\,\mathrm{Tr}R^3-5\,\mathrm{Tr}R\Big).
\end{equation}
At an $\mathcal{N}=1$ SCFT, the correct $U(1)_R$ is singled out by $a$-maximization \cite{intriligatorExactSuperconformalRSymmetry2003}: among all anomaly-free trial R-symmetries, the physical one is the choice that maximizes $a$.

Once a candidate R-symmetry is obtained, consistency with RG flow is checked using the $a$-theorem \cite{komargodskiRenormalizationGroupFlows2011}, which states that along any unitary RG flow from a UV fixed point to an IR fixed point one has
\begin{equation}
a_{\mathrm{UV}} \ge a_{\mathrm{IR}},
\end{equation}
with equality only at a fixed point. In practice, this provides a nontrivial check that the putative IR SCFT data does not violate basic monotonicity.

A further necessary condition is that the gauge coupling reach a conformal fixed point. In an $\mathcal{N}=1$ gauge theory, the exact NSVZ beta function takes the form \cite{novikovExactGellMannLowFunction1983}
\begin{equation}
\beta(g)= -\frac{g^3}{16\pi^2}\,
\frac{3T(G)-\sum_i T(R_i)\big(1+\gamma_i\big)}{1-\frac{g^2}{8\pi^2}T(G)}\,,
\end{equation}
where $T(G)$ is the Dynkin index of the adjoint, $T(R_i)$ that of the matter representation $R_i$, and $\gamma_i$ the anomalous dimension of the corresponding chiral superfield. At a fixed point, the numerator must vanish:
\begin{equation}
3T(G)-\sum_i T(R_i)\big(1+\gamma_i\big)=0.
\end{equation}
In a SCFT, the anomalous dimensions are related to R-charges via $\Delta=\frac{3}{2}R$ and $\Delta=1-\frac{1}{2}\gamma$, so for each chiral field one may equivalently write\footnote{Note that sometimes the NSVZ beta function is written with the opposite sign convention for the anomalous dimensions. Here I use $\gamma_i \equiv \dod{\ln Z_i}{\ln\mu}$, as is the typical convention when writing the AMSB soft terms as in \cref{eq:masses+trilinear}.} $\gamma_i=2-3R_i$. Thus, NSVZ beta function provides a constraint that the R-charges selected by $a$-maximization must satisfy.

Because the operator spectrum must remain unitary, one must also consider possible decoupling. For a chiral primary operator in a unitary $\mathcal{N}=1$ SCFT, the unitarity bound implies $R\ge 2/3$ (equivalently $\Delta\ge 1$). If a gauge-invariant chiral operator violates this bound in the naive analysis, it becomes free and decouples, potentially introducing an emergent symmetry that can affect the correct R-symmetry and the stress-tensor anomaly coefficients. Therefore, in each case below I scan the gauge-invariant chiral spectrum for operators that would violate the unitarity bound; whenever such decoupling is triggered, I include the contribution of the resulting free singlet(s) to $a$ and $c$ in the IR central charges.

Finally, I check the conformal collider bounds \cite{hofmanConformalColliderPhysics2008}, which constrain the stress-tensor three-point function coefficients in any 4d CFT to
\begin{equation}
\frac{1}{2}\le \frac{a}{c}\le \frac{3}{2}.
\end{equation}
This serves as an additional, genuinely dynamical constraint on the candidate IR fixed point.

In general it is also important to keep track of (potentially) allowed deformations. In $\mathcal{N}=1$ theories, an $R$-symmetry selection rule implies that a gauge-invariant chiral operator $\mathcal{O}$ can appear as a superpotential deformation $\delta W\sim \mathcal{O}$ only if $R(\mathcal{O})=2$ (marginal) or $R(\mathcal{O})<2$ (relevant). Although symmetry and $R$-charge considerations could, in principle, allow such deformations, in the present analysis the UV superpotential is taken to be identically zero and I have already ruled out dynamically generated superpotentials from the outset. The question of deformations is then not of relevance here.

\FloatBarrier
\subsection{$Sp(10)$ with $10_F+110_{A3}$}
Let $Q$ denote the $\ydiagram{1}$ and $A$ denote the $\ydiagram{1,1,1}$. The UV theory has global symmetries $U(1)_Q\times U(1)_A\times U(1)_R$ where $U(1)_Q$ and $U(1)_A$ rotate the phase of $Q$ and $A$, respectively, and $U(1)_R$ is the R-symmetry. These are anomalous, however, and the surviving non-anomalous combinations are given by $U(1)_X\times U(1)_R$ where
\begin{align}
    R(Q) + 27 R(A) =& 16 \\
    X(Q) = 27,\qquad X(A) &= -1
\end{align}
At the putative fixed point $\gamma_A = 2-3R(A),$ and $\gamma_Q = 2-3R(Q)$, such that the NSVZ beta function is proportional to
\begin{align}
    3T(G) - T(A)(1+\gamma_A) - T(Q)(1+\gamma_Q) =& 3\times 6 - \frac{27}{2}(3-3R(A)) - \frac{1}{2}(3-3R(Q)) \\
    =& 18 - 3\frac{28}{2} + \frac{3}{2}(R(Q)+27R(A)) 
    \\
    =& -24 + \frac{3}{2}(16) = 0,
\end{align}
so any choice of $U(1)_R$ in this continuous family would be consistent with a fixed point. Because there is this continuous family of consistent $U(1)_R$'s, we find the one that maximizes the central charge $a$. Let $R(A)=x$ so that $R(Q)=16-27x$. Then
\begin{align}
    a(x) =& \frac{3}{32}(\text{Tr}(R^3) - \text{Tr}(R)) = \frac{3}{32}\Big(100990 -545600\,x +983160\,x^2 -590160\,x^3\Big).
\end{align}
This is maximized for $x=\frac{\sqrt{43729}+8193}{14754}\approx 0.56948$, with $a_{\rm IR}\approx 12.4866$. We then have that at the putative fixed point 
\begin{align}
    R(Q) =& \frac{4951-9 \sqrt{43729}}{4918}\approx 0.624027,
    &
    R(A) =& \frac{\sqrt{43729}+8193}{14754}\approx 0.56948.
\end{align}
Note that the lowest dimensional gauge invariants are of the form $Q^2A^2$ and $A^4$, which would have $R\approx 2.38702,\, 2.27792$, respectively, neither of which violate unitarity. So we do not have to worry about any decoupling. Moreover, since there are no gauge invariants with $R\leq 2$ there are no relevant or marginal deformations possible.

At the putative fixed point the central charges would be
\begin{align}
    a_{\rm IR} =& \frac{5 \left(87458 \sqrt{43729}+706532805\right)}{290240688}\approx 12.4866,
    &
    c_{\rm IR} =& \frac{5 \left(63401 \sqrt{43729}+342108690\right)}{145120344}\approx 12.2439.
\end{align}
This is compared to the UV free theory in which $a_{\rm UV} = \frac{205}{16}= 12.8125$, so $a_{\rm IR}<a_{\rm UV}$ in accordance with the $a$-theorem. We also have
\begin{align}
    \frac{1}{2}< \frac{a_{\rm IR}}{c_{\rm IR}}\approx 1.01982 < \frac{3}{2},
\end{align}
consistent with conformal collider bounds. 

In summary I find no evidence against this theory flowing to a SCFT in the IR. This is by no means a proof, and as I noted earlier AMSB could deflect the RG flow away from such a fixed point anyway. 

\FloatBarrier
\subsection{$Sp(4)$ with $16_{(1,1)}$}
Let $Q$ be the 16. I find that the only anomaly free R-charge is $R(Q)=4/19$. At the putative fixed point then $\gamma_Q = 2-3R(Q) = 26/19$, which does lead to a vanishing NSVZ beta function $3 T_G - T(Q)(1+\gamma_Q) = 3\times 3 - \frac{19}{5}(1+\frac{26}{19})=0$. In order for a gauge invariant of type $Q^k$ to decouple, we would need $R(Q^k) < \frac{2}{3}$ which is only possible for $k=2,3$. There is no singlet in $\text{Sym}^2(16)$ or $\text{Sym}^3(16)$. So there is no decoupling. At the putative fixed point the central charges would be
\begin{align}
    a_{\rm IR}=&\frac{46365}{54872}
\approx 0.8450
,
    &
    c_{\text{IR}}=&\frac{27695}{27436}
\approx 1.0094,
\end{align}
compared to the UV value of 
\begin{align}
    a_{\rm UV}=\frac{53}{24}\approx 2.2083.
\end{align}
So $a_{\rm IR} < a_{\rm UV}$ and the $a$-theorem is satisfied. We also have
\begin{align}
    \frac{1}{2}< \frac{a_{\rm IR}}{c_{\rm IR}} \approx 0.8371 < \frac{3}{2}
\end{align}
consistent with conformal collider bounds. 

In summary I find no evidence against this theory flowing to a SCFT in the IR. This is by no means a proof, and as I noted earlier AMSB could deflect the RG flow away from such a fixed point anyway.

\subsection{$Sp(8)$ with $48_{A3}$}
Let $A$ denote the $\ydiagram{1,1,1}$. Since there is only one chiral superfield, one readily arrives at only one possible non-anomalous $U(1)_R$ with $R(A) = 2/7$. At the putative fixed point then $\gamma_A = 2-3R(A) = 8/7$. The NSVZ beta function does indeed vanish here as $3T_G - T(A)(1+\gamma_A) = 3\times 5 - 7(1+\frac{8}{7}) = 0$. A chiral primary of the form $A^k$ could only violate unitarity for $k=2$, but there is no singlet in $\text{Sym}^2(48)$, so we need not worry about any decoupling. The central charges would be
\begin{align}
    a_{\rm IR} =& \frac{6921}{1372}\approx 5.04, 
    &
    c_{\rm IR} =&\frac{3387}{686}\approx 4.94,
\end{align}
whereas in the UV 
\begin{align}
    a_{\rm UV} =& \frac{31}{4} = 7.75.
\end{align}
So $a_{\rm IR} < a_{\rm UV}$ and the $a$-theorem is satisfied. We also have
\begin{align}
\frac{1}{2} < \frac{a_{\rm IR}}{c_{\rm IR}}\approx 1.02 
< \frac{3}{2}
\end{align}
consistent with conformal collider bounds. 

In summary I find no evidence against this theory flowing to a SCFT in the IR. This is by no means a proof, and as I noted earlier AMSB could deflect the RG flow away from such a fixed point anyway.

\subsection{$Sp(12)$ with $208_{A3}$}

Let $A$ denote the $\ydiagram{1,1,1}$. Since there is only one chiral superfield, one readily arrives at only one possible non-anomalous $U(1)_R$. I find that $R(A) = 15/22$. At the putative fixed point then $\gamma_A = 2-3R(A) = -1/22$. This gives an NSVZ beta function proportional to:
\begin{align}
    3T_G - T(A)(1+\gamma_A) =& 3\times 7 - 22(1-1/22) = 21-21 = 0,
\end{align}
so there is no obvious issue with this theory flowing to a SCFT in the IR. At the putative fixed point the central charges would be:
\begin{align}
    a_{\rm IR}=&\frac{403455}{21296}\approx 18.945, & c_{\rm IR}=& \frac{387725}{21296}\approx 18.207.
\end{align}
In the UV free theory we would instead have $a_{\text{UV}}=\frac{455}{24}\approx 18.958 > a_{\rm IR}$, consistent with the $a$-theorem. Note that apparent unitarity violation of some composite $\sim A^k$ would require $k < \frac{44}{45}$ which is not possible, so there is no worry about decoupling. We also have
\begin{align}
    \frac{1}{2} < \frac{a_{\rm IR}}{c_{\rm IR}}\approx 1.04
    < \frac{3}{2}
\end{align}
consistent with conformal collider bounds.

In total, then, I find no evidence against this theory flowing to a SCFT in the IR. This is by no means a proof, and as I noted earlier AMSB could deflect the RG flow away from such a fixed point anyway.

\section{Reconciling with Tumbling?}\label{sec:ReconcileTumbling}
Ref~\cite{cacciapaglia_perusing_2026} arrives at their conclusions through a perfectly standard application of the tumbling hypothesis, and I agree with all of their conclusions within that framework. The AMSB framework detailed in this work, however, starkly disagrees. 

For the tumbling approach used in \cite{cacciapaglia_perusing_2026}, one considers scalar bilinear operators built from the symmetric part of two fermion reps,
$(\psi_r\psi_r)_R$ with $R\subset {\rm Sym}(r\otimes r)$.
The ``attractiveness" is estimated by
\begin{equation}
    \Delta \;=\; 2\,C_r \;-\; C_R,
\end{equation}
where $C_r$ is the quadratic Casimir of the fermion representation\footnote{When the fermions are of different types this generalizes to $C_{r_1}+C_{r_2}$.} $r$ and $C_R$ is the quadratic Casimir of the channel representation $R$. When $\Delta < 0$ the channel is considered ``repulsive" and condensation is assumed to be impossible. When $\Delta >0$ the channel is ``attractive" and such channels are ranked as more likely to condense the larger $\Delta $ is, with the ``Most Attractive Channel" (MAC) being the most likely. 

To organize this comparison, I consider increasingly nonstandard departures from conventional tumbling in the following order: first, an attractive NMAC bilinear whose VEV lies on a non-maximal orbit; second, a bilinear in a channel classified as repulsive by the usual Casimir criterion; and finally, a higher-order fermion condensate. For each theory I stop once I identify a channel capable of reproducing the AMSB stabilizer. Thus, the analysis is not intended as an exhaustive classification of all higher-order condensates in cases where a suitable bilinear has already been found.

For example, consider $\mathcal{N}=0$ $SU(6)$ with a single $20_{A3}$. As in \cite{cacciapaglia_perusing_2026}, the only attractive channel is the adjoint. If we restrict to maximal regular subgroups, we find the same three tumbling options: (A) $SU(6)\to SU(5)\times U(1)$, (B) $SU(6)\to SU(4)\times SU(2)\times U(1)$ and (C) $SU(6)\to SU(3)\times SU(3)\times U(1)$. If one instead allows the \textit{repulsive} channel $\mathbf{175}$ to condense, we can realize breaking to just $SU(3)\times SU(3)$. For example, take:
\begin{align}
    \braket{A^{abc}A^{def}} =& \begin{cases}
        v \epsilon^{abc}\epsilon^{def} & \{a,b,c,d,e,f\}\in\{1,2,3\} \\
        0 & \text{ else}
    \end{cases}
\end{align}
In the first three theories considered above, the same sort of thing happens: the only tumbling channels that could realize the AMSB symmetry breaking pattern are repulsive. For the fourth theory, Spin(12) with a $\mathbf{32}$, there are no bilinear condensates that can break to the AMSB stabilizer, and the first such possibility for derivative-free channels occurs at the six-fermion order. For the last two theories, I find that one can break to the AMSB stabilizer in the NMAC if one allows for VEVs not considered in \cite{cacciapaglia_perusing_2026}. I list all six cases in \cref{tab:am sb-tumbling-stabilizers} and discuss the details of the group theory in \cref{app:groupDetails}. The first three cases are quite radical: while non-maximal tumbling is not necessarily inconceivable, the conventional wisdom is that condensates should not be able to form in repulsive channels. While that conclusion is not a rigorous theorem, it seems quite reasonable. It is possible that the notion of bilinear condensates, while applicable to QCD, is not the right way to think about confinement in general. The notion of tumbling through a six-fermion channel in the fourth theory is likewise not conventionally considered a possibility. 

\begin{table}[t]
\centering
\renewcommand{\arraystretch}{1.8}
\begin{tabular}{|l|l|l|l|}
\hline
\textbf{Theory} &
\textbf{Tumbling stabilizer} &
\textbf{AMSB stabilizer} &
\textbf{Channel for AMSB} \\
\hline
\makecell[l]{SU(6) with $20_{A3}$} &
\makecell[l]{$SU(5)\times U(1)$\\ $SU(4)\times SU(2)\times U(1)$\\ $SU(3)\times SU(3)\times U(1)$} &
\makecell[l]{$SU(3)\times SU(3)$} &
\makecell[l]{$R=\mathbf{175}$ (repulsive)} \\
\hline
\makecell[l]{$E_7$ with $56_F$} &
\makecell[l]{$E_6\times U(1)$} &
\makecell[l]{$E_6$} &
\makecell[l]{$R=\mathbf{1463}$ (repulsive)} \\
\hline
\makecell[l]{Spin(11) with $32_{\rm spin}$} &
\makecell[l]{$\text{Spin}(10)$\\ $\text{Spin}(9)\times U(1)$} &
\makecell[l]{$SU(5)$} &
\makecell[l]{$R=\mathbf{462}$ (repulsive)} \\
\hline
\makecell[l]{Spin(12) with $32_{\rm spin}$} &
\makecell[l]{$SU(6)\times U(1)$\\ $\text{Spin}(10)\times U(1)$} &
\makecell[l]{$SU(6)$} &
\makecell[l]{$\psi^6$ condensate} \\
\hline
\makecell[l]{Spin(13) with $64_{\rm spin}$} &
\makecell[l]{$\text{Spin}(11)\times U(1)$\\ $\text{Spin}(10)\times SU(2)$} &
\makecell[l]{$SU(3)\times SU(3)$} &
\makecell[l]{$R=\mathbf{286}$ \\ (NMAC, non-maximal) } \\
\hline
\makecell[l]{Sp(6) with $6_F+14'_{A_3}$} &
\makecell[l]{$SU(3)\times U(1)$\\ $SU(2)\times Sp(4)$\\ $SU(2)\times U(1)\times U(1)$} &
\makecell[l]{$SU(2)$} &
\makecell[l]{
$R=\mathbf{14}$ \\ (NMAC, non-maximal)} \\
\hline
\end{tabular}
\caption{Tumbling versus AMSB stabilizers for the six pseudoreal theories, and the tumbling channels required to reproduce the AMSB stabilizer. The first three require tumbling through repulsive channels. The fourth cannot reproduce the AMSB stabilizer with bilinear condensates (but could at sixth-order). The last two can reproduce the AMSB stabilizers by tumbling through the NMAC via non-maximal directions.}
\label{tab:am sb-tumbling-stabilizers}
\end{table}

Regarding the first three theories, we appear to be forced into one of two possibilities: (i) the AMSB conjecture fails in this collection of theories, or (ii) condensates can form in repulsive channels. I do not intend to argue one way or the other in this work. The ultimate conclusion requires non-perturbative methods applied directly to the theory of interest, such as lattice simulations.

I do want to point out, however, that this is much stronger than the disagreements encountered in earlier AMSB–tumbling comparisons. In particular, in the analyses of chiral theories in \cite{csaki_exact_2021,csaki_more_2022}, AMSB could be reconciled with the standard tumbling framework only after allowing condensation in the next-most-attractive channel (NMAC) in addition to the MAC; importantly, these channels are still \emph{attractive} in the usual tumbling ranking. Similarly, in \cite{leedom_exact_2025}, the apparent “tensions” were resolved by taking seriously the possibility of competition among multiple attractive channels, without needing any modification of the basic attractiveness/repulsiveness criterion.  

By contrast, reproducing the AMSB stabilizer subgroups in three of the present pseudoreal class requires condensation precisely in those channels that the standard tumbling estimate classifies as \emph{repulsive} (i.e. with $\Delta<0$). While the tumbling hypothesis is of course heuristic and the attractiveness ranking can be interpreted as a proxy for “likelihood to condense,” it is comparatively uncontroversial to imagine that several attractive channels might compete or that the realized breaking pattern depends on parameters. But requiring that repulsive channels should condense would be a qualitatively more radical dynamical input.

If interpreted as the AMSB conjecture failing in this class of theories, that is interesting as well. To date the explicit examples of phase transitions at $m\sim\Lambda$ consist solely of theories with a classically conformal superpotential \cite{baiPhasesConfiningSU52022,delimaSconfiningSUSYQCDAnomaly2023}. In this work all six examples are ADS-like instead. One as biased to the AMSB conjecture as myself could take this as an argument against the tumbling result, but one could just as well take this as an indication that the AMSB conjecture fails for a wider array of theories than previously indicated.

\section{Conclusion}\label{sec:Conclusion}

In \cite{cacciapaglia_perusing_2026}, the authors proposed, using the tumbling hypothesis together with anomaly and symmetry constraints, that most confining pseudoreal gauge theories outside the conformal window should exhibit at least one massless spin-1 state in the deep IR. In this paper I revisit this conclusion by analyzing the corresponding asymptotically free $\mathcal N=1$ pseudoreal theories perturbed by AMSB, and by taking seriously the conjecture that the AMSB and non-SUSY dynamics lie in the same universality class.

The main result is that for the six $\mathcal N=1$ theories whose non-SUSY limits were identified in \cite{cacciapaglia_perusing_2026} as non-conformal, the AMSB vacua never leave an isolated $U(1)$ factor in the residual gauge symmetry. In all cases, the UV gauge bosons are either Higgsed into massive vector multiplets or belong to confining non-abelian factors, and no spin-1 composites remain massless. Moreover, in all but one case the non-SUSY universality class predicted to emerge from these AMSB vacua is \emph{gapped} and does not contain any massless fields at all. These conclusions directly contradict the tumbling-based expectations of \cite{cacciapaglia_perusing_2026} for the same set of theories.

I then ask whether this tension can be resolved without abandoning the AMSB universality conjecture. The answer is that matching the AMSB symmetry breaking patterns within a tumbling framework would require condensation in \emph{repulsive} tumbling channels in three cases, and in one case requires tumbling through a six-fermion channel. This is much more radical than the usual ``competition among attractive channels'' logic, and it suggests that either (i) the universality-class identification underlying the AMSB cross-check is more subtle than assumed for this class, or (ii) the tumbling picture needs to be generalized beyond the standard categorization of channels by attractiveness.

For completeness, I also investigated the four additional pseudoreal theories whose non-SUSY limits were listed as potentially within the conformal window. On the supersymmetric side, none of these theories admits the ``regular'' confining/ADS-like mechanisms ruled out by the $\iota$-index, and each admits an anomaly-free $U(1)_R$ consistent with the NSVZ fixed-point condition and with basic unitarity/central-charge constraints. This provides evidence compatible with an interacting IR SCFT in the SUSY theories, though (as emphasized in the main text) AMSB could deflect the RG flow away from such fixed points, so these results do not decisively determine the conformal/non-conformal character of the corresponding non-SUSY limits.

Overall, the conclusions are clear: the spectrum expected from tumbling is not reproduced by the AMSB analysis under the universality conjecture, and any reconciliation through tumbling would require qualitatively non-standard dynamical input. Deciding between these possibilities requires genuinely non-perturbative methods applied to the non-SUSY pseudoreal theories themselves (e.g.\ lattice simulations and/or functional RG). 

My results identify an interesting and drastic tension between two preeminent means of analyzing strongly coupled gauge theories. The ultimate resolution of this tension could have important ramifications for either: limiting the scope of the AMSB conjecture or reducing the predictivity of the tumbling hypothesis through the expansion of potential condensation channels.

\acknowledgments
I would like to thank Hitoshi Murayama, Riku Ishikawa, and Shota Saito who obtained the same AMSB deformation of the t-confining pseudoreal theory $SU(6)+\ydiagram{1,1,1}$ independently, and were kind enough to exchange drafts and compare results. I thank Hitoshi Murayama as well for his comments on this draft. I thank Clara Xu for her help with deciding on a title. Giacomo Cacciapaglia and Konstantinos Kollias of \cite{cacciapaglia_perusing_2026} provided a helpful reading of the draft, catching several typos and mistakes, for which I am grateful. Finally, I thank 
\`Alvaro Pastor Guti\'errez for finding a typo and bringing the works \cite{liConfinementSymmetryBreaking2026,liDynamicalSymmetryBreaking2025} on functional RG to my attention.

\appendix
\crefalias{section}{appsec}
\section{Tumbling to the AMSB Stabilizer}\label{app:groupDetails}
Given a condensate in a channel $R$, a VEV $v\in R$ breaks the connected gauge symmetry to the stabilizer
$\mathfrak{stab}(v)=\{X\in\mathfrak{g}\,|\,X\cdot v=0\}$.
To test whether one can end with a purely non-abelian stabilizer $H$ (and no residual isolated $U(1)$’s), we choose a maximal embedding
\[
G\supset H\times U(1)^k,
\qquad
R\downarrow_{H\times U(1)^k}=\bigoplus_i \big(r_i\otimes q^{(i)}\big),
\]
and restrict to the $H$-singlet components ($r_i=\mathbf{1}$), since any non-singlet VEV would break $H$.
For the abelian factors, write the $U(1)^k$ generators as $Q_a$ and an arbitrary $Q(a)=\sum_a \alpha_a Q_a$.
If the VEV uses $H$-singlets with charge vectors $\{q^{(i)}\}$, then $Q(a)$ remains unbroken iff
$\alpha\cdot q^{(i)}=0$ for every charge direction present in the VEV.
Hence, $U(1)^k$ is completely Higgsed precisely when the charge vectors used by the VEV span $\mathbb{R}^k$.
Reality constraints (real/pseudoreal $R$) only restrict which charge directions can appear with independent coefficients, but the connected residual $U(1)$’s are still determined by the same condition.

In all but one case I restrict my attention to bilinear condensates. However, in the $\text{Spin}(12)$ theory no such bilinear condensate exists that could realize the AMSB stabilizer, and so I consider condensates with higher numbers of fermions. It is of course possible that higher-order condensates play a role in the real dynamics of any of these theories, but it is only in this case that the AMSB result forces us to consider them.

In the remainder of this appendix I work out the group theory details for my claims about which channels could realize the AMSB stabilizer subgroups. I use \texttt{LieART} \cite{fegerLieART20Mathematica2020} and \texttt{GroupMath} \cite{fonsecaGroupMathMathematicaPackage2021} for the decompositions. 

\subsection{$SU(6)+\ydiagram{1,1,1}$}
Note that $\text{Sym}^2(\mathbf{20}) = \mathbf{35}+\mathbf{175}$. The $\mathbf{35}$ is the adjoint and so always leaves an isolated $U(1)$. So the only bilinear channel left to consider is the $\mathbf{175}$.

Consider the maximal embedding
\begin{align}
SU(6)\supset SU(3)_A\times SU(3)_B\times U(1)_X .
\end{align}
The $\mathbf{175}$ branches as
\begin{align}
\mathbf{175}\downarrow_{SU(3)\times SU(3)\times U(1)}
\supset (\mathbf{1},\mathbf{1})_6+(\mathbf{1},\mathbf{1})_0+(\mathbf{1},\mathbf{1})_{-6}+\cdots .
\end{align}
We can choose $\braket{\mathbf{175}}\in(\mathbf{1},\mathbf{1})_6$, for instance. This preserves the $SU(3)\times SU(3)$ but breaks the remaining $U(1)$. For this maximal embedding, there are no other generators commuting with the $SU(3)\times SU(3)$,
so no other isolated abelian factor can remain.

So we indeed can have a VEV for the $\mathbf{175}$ such that
\begin{align}
    SU(6)\to& SU(3)\times SU(3)
\end{align}
with no residual $U(1)$.
 
Note that the $\mathbf{20}$ decomposes under $SU(3)\times SU(3)$ as
\begin{align}
    \mathbf{20}\downarrow_{SU(3)\times SU(3)} =& 2 (\mathbf{1},\mathbf{1}) + (\mathbf{3},\overline{\mathbf{3}}) + (\overline{\mathbf{3}},\mathbf{3}).
\end{align}
Since the massless content of the residual $SU(3)\times SU(3)$ is made up of singlets and vectorlike fermions, there is no further gauge-breaking.

\subsection{$E_7+\ydiagram{1}$}
Note that $\text{Sym}^2(\mathbf{56}) = \mathbf{133}+\mathbf{1463}$. Since $\mathbf{133}$ is the adjoint, the only bilinear to consider is the $\mathbf{1463}$.

Consider the maximal embedding
\begin{align}
    E_7\supset& E_6\times U(1)
\end{align}
The $\mathbf{1463}$ branches as
\begin{align}
    \mathbf{1463}\downarrow_{E_6\times U(1)} = 
    \mathbf{1}_6+\mathbf{1}_0+\mathbf{1}_{-6} + \cdots.
\end{align}
We can choose $\braket{\mathbf{1463}}=\alpha \mathbf{1}_{+ 6} + \alpha^* \mathbf{1}_{-6}$ (note that $\mathbf{1463}$ is strictly real). The stabilizer will contain $E_6$ but not the remaining $U(1)$. Because the embedding is maximal this precludes any larger stabilizer, so 
\begin{align}
    E_7 \to E_6
\end{align}
with no residual $U(1)$. Under the residual we have
\begin{align}
    \mathbf{56}\downarrow_{E_6} =& \mathbf{1}+\mathbf{1}+\mathbf{27}+\overline{\mathbf{27}}
\end{align}
Since the massless matter consists of singlets and vectorlike pairs, there is no further gauge breaking.

\subsection{Spin(11) $+32_{\rm spin}$}
Note that $\text{Sym}^2(\mathbf{32}) = \mathbf{11}+\mathbf{55}+\mathbf{462}$. The $\mathbf{55}$ is the adjoint and so can only break $\text{Spin}(11)\to \text{Spin}(9)\times U(1)$. The $\mathbf{11}$ is the fundamental and so can only break $\text{Spin}(11)\to \text{Spin}(10)$. The only remaining bilinear channel to consider is the $\mathbf{462}$.

Consider the maximal embedding 
\begin{align}
    \text{Spin}(11)\supset& SU(5)\times U(1)
\end{align}
The $\mathbf{462}$ branches as
\begin{align}
    \mathbf{462}\downarrow_{SU(5)\times U(1)} =& \mathbf{1}_{10} + \mathbf{1}_0 + \mathbf{1}_{-10} + \cdots .
\end{align}
We can choose the VEV along $\mathbf{1}_{ 10} + \mathbf{1}_{-10}$ (the $\mathbf{462}$ is strictly real) which will contain $SU(5)$ as its stabilizer but not the $U(1)$. Since the embedding is maximal, there is no larger stabilizer and 
\begin{align}
    \text{Spin}(11)\to& SU(5)
\end{align}
with no residual $U(1)$. Under the unbroken $SU(5)$ the massless matter decomposes as
\begin{align}
    \mathbf{32}\downarrow_{SU(5)} =& \mathbf{1}+\mathbf{1}+\mathbf{5}+\overline{\mathbf{5}}+\mathbf{10}+\overline{\mathbf{10}}.
\end{align}
Since the massless fermions are singlets and vectorlike pairs, there is no further gauge breaking.

\subsection{Spin(12)$+32_{\rm spin}$}
Note that $\text{Sym}^2(\mathbf{32}_-)=\mathbf{66}+\mathbf{462}_-$. The $\mathbf{66}$ is the adjoint and so always leaves an isolated $U(1)$. The only bilinear to consider is then the repulsive $\mathbf{462}_-$.

Under the maximal subgroup
\begin{align}
    SU(6)\times U(1)\subset& \text{ Spin}(12)
\end{align}
the $\mathbf{462}_-$ decomposes as
\begin{align}
    \mathbf{462}_-\downarrow_{SU(6)\times U(1)} =& \mathbf{21}_{-4}+\overline{\mathbf{21}}_{4} + \mathbf{35}_0 + \mathbf{105}_2 + \overline{\mathbf{105}}_{-2} + \mathbf{175}_0 .
\end{align}
This contains no $SU(6)$ singlets so it is not possible for a $\mathbf{462}_-$ VEV to break $\text{Spin}(12)\to SU(6)$. 

In order to reproduce the AMSB stabilizer we are then forced to consider higher-order condensates. Decomposing the allowed four-fermion scalar sector gives\footnote{Here $\mathbb{S}_\lambda$ is the Schur functor on the partition $\lambda$. Spin statistics and Lorentz invariance require the derivative-free $2n$-fermion composites of irrep $R$ to live in $\mathbb{S}_{\underbrace{(2,...,2)}_{n}}(R)$.} 
\begin{align}
\mathbb S_{(2,2)}(\mathbf{32}_-)
\downarrow_{SU(6)\times U(1)}
\supset5\,\mathbf1_0,
\end{align}
with no other $SU(6)$ singlets. Consequently, no local derivative-free spin-zero four-fermion condensate can preserve $SU(6)$ while Higgsing the commuting $U(1)$. At sixth order, however,
\begin{align}
 \mathbb{S}_{(2,2,2)}(\mathbf{32}_-)
 \downarrow_{SU(6)\times U(1)}
 \supset3\,\mathbf 1_{-6}\oplus 10\,\mathbf 1_0
 \oplus 3\,\mathbf 1_{+6}.
\end{align}
Here the chirality labels are defined by
\begin{align}
 \mathbf{32}_-
 &\equiv \Lambda^{\mathrm{odd}}\mathbf{6}
 \downarrow_{SU(6)\times U(1)}
 =\mathbf{6}_{-2}\oplus\mathbf{20}_0\oplus\overline{\mathbf{6}}_{+2},
 \\
\mathbf{32}_+
 &\equiv \Lambda^{\mathrm{even}}\mathbf{6}
 \downarrow_{SU(6)\times U(1)}
 =\mathbf 1_{-3}\oplus\mathbf{15}_{-1}
  \oplus\overline{\mathbf{15}}_{+1}\oplus\mathbf 1_{+3},
\end{align}
up to an overall reversal of the $U(1)$ charges.  Correspondingly,
there are two inequivalent $\text{Spin}(12)$ representations of dimension
$462$, which we denote by
\begin{align}
 \mathbf{462}_-
 &\equiv (0,0,0,0,2,0)
 \subset \operatorname{Sym}^2(\mathbf{32}_-),
 \\
\mathbf{462}_+
 &\equiv (0,0,0,0,0,2)
 \subset \operatorname{Sym}^2(\mathbf{32}_+).
\end{align}
The bilinear of the fermion in the present theory therefore contains
$462_-$,
\begin{align}
 \operatorname{Sym}^2(\mathbf{32}_-)=\mathbf{66}\oplus\mathbf{462}_-,
\end{align}
and $\mathbf{462}_-$ contains no $SU(6)$ singlet.  The charged singlets at
sixth order instead arise from the opposite-chirality representation
$\mathbf{462}_+$ and from
\begin{align}
 \mathbf{84942}_+\equiv(0,0,0,1,0,2).
\end{align}
More precisely, the full $\text{Spin}(12)$ decomposition contains
\begin{align}
 \mathbb{S}_{(2,2,2)}(\mathbf{32}_-)
 \supset
 2\,\mathbf{462}_+\oplus\mathbf{84942}_+\oplus\cdots ,
\end{align}
where
\begin{align}
 \mathbf{462}_+
 \downarrow_{SU(6)\times U(1)}
 &\supset
 \mathbf 1_{-6}\oplus\mathbf 1_0\oplus\mathbf 1_{+6},
 \\
 \mathbf{84942}_+
 \downarrow_{SU(6)\times U(1)}
 &\supset
 \mathbf 1_{-6}\oplus\mathbf 1_0\oplus\mathbf 1_{+6}.
\end{align}
Thus, the two copies of $462_+$ provide two of the three charged-singlet
pairs, while $84942_+$ provides the third.  A real VEV involving a
conjugate pair,
\begin{align}
 \langle\mathcal O_{+6}\rangle
 =\langle\mathcal O_{-6}\rangle^*,
\end{align}
preserves $SU(6)$ while Higgsing the commuting continuous $U(1)$ and
can therefore reproduce the connected AMSB stabilizer.

One may also define a heuristic extension of the MAC diagnostic by
assuming that the six-body interaction is a sum of equal pairwise
gauge-exchange kernels.  On a state in a total $\operatorname{Spin}(12)$
representation $X$,
\begin{align}
 \sum_{i<j}T_i^aT_j^a
 =\frac12\left[C_2(X)-6C_2(32_-)\right],
\end{align}
so it is natural to introduce
\begin{align}
 \Delta_X^{(6)}
 \equiv 6C_2(32_-)-C_2(X).
\end{align}
Using
\begin{align}
 C_2(\mathbf{32}_-)=\frac{33}{4},\qquad
 C_2(\mathbf{462}_+)=18,\qquad
 C_2(\mathbf{84942}_+)=38,
\end{align}
gives
\begin{align}
 \Delta_{\mathbf{462}_+}^{(6)}
 &=\frac{63}{2},
 &\sum_{i<j}T_i^aT_j^a&=-\frac{63}{4},
 \\
 \Delta_{\mathbf{84942}_+}^{(6)}
 &=\frac{23}{2},
 &\sum_{i<j}T_i^aT_j^a&=-\frac{23}{4}.
\end{align}
Both channels are therefore net attractive in this equal-pairwise
Casimir approximation, with $\mathbf{462}_+$ the more attractive of the two.
This should not be confused with the ordinary bilinear MAC criterion. Moreover, the generalized criterion does not determine the
complete six-body kernel and cannot resolve mixing between the two
copies of $\mathbf{462}_+$.  The calculation therefore establishes that the
required channel is kinematically allowed and heuristically attractive,
but not that the corresponding condensate necessarily forms.

\subsection{Spin(13)$+64_{\rm spin}$}
Note that $\text{Sym}^2(\mathbf{64})=\mathbf{78}+\mathbf{286}+\mathbf{1716}$. The $\mathbf{78}$ is the adjoint and can only break $\text{Spin}(13)\to \text{Spin}(11)\times U(1)$. This leaves the only bilinear options as the $\mathbf{286}$ which is the NMAC, and the $\mathbf{1716}$ which is repulsive.

A maximal subgroup is
\begin{align}
    \text{Spin}(13)\supset& SU(3)\times SU(3)\times U(1)\times U(1)
\end{align}
The $\mathbf{1716}$ and $\mathbf{286}$ decompose as
\begin{align}
    \mathbf{1716}\downarrow_{SU(3)\times SU(3)\times U(1)\times U(1)} =& (\mathbf{1},\mathbf{1})_{6,0} + (\mathbf{1},\mathbf{1})_{-6,0} + (\mathbf{1},\mathbf{1})_{0,6} + (\mathbf{1},\mathbf{1})_{0,-6} + \cdots,
    \\
    \mathbf{286}\downarrow_{SU(3)\times SU(3)\times U(1)\times U(1)} =& (\mathbf{1},\mathbf{1})_{3,3}+(\mathbf{1},\mathbf{1})_{3,-3} + (\mathbf{1},\mathbf{1})_{-3,3} + (\mathbf{1},\mathbf{1})_{-3,-3} + \cdots.
\end{align}
In either case is possible to choose a VEV that is stabilized by $SU(3)\times SU(3)$ but breaks the remaining $U(1)\times U(1)$. For example (keeping in mind that the $\mathbf{286}$ is real) one could take
\begin{align}
    \braket{\mathbf{286}} =& \alpha (\mathbf{1},\mathbf{1})_{3,3}+\beta(\mathbf{1},\mathbf{1})_{3,-3} + h.c. 
\end{align}
with $\alpha\neq \beta$ and $\alpha,\beta\neq 0$. 

There are no remaining commuting generators available, so no larger stabilizer possible and we have
\begin{align}
    \text{Spin}(13)\to& SU(3)\times SU(3)
\end{align}
with no residual $U(1)$. Note that the massless fermions decompose as
\begin{align}
    \mathbf{64}\downarrow_{SU(3)\times SU(3)} =& 4(\mathbf{1},\mathbf{1})
+2(\mathbf{3},\mathbf{1})
+2(\mathbf{1},\mathbf{3})
+2(\overline{\mathbf{3}},\mathbf{1})
+2(\mathbf{1},\overline{\mathbf{3}})
+(\mathbf{3},\mathbf{3})
+(\mathbf{3},\overline{\mathbf{3}})
+(\overline{\mathbf{3}},\mathbf{3})
+(\overline{\mathbf{3}},\overline{\mathbf{3}}).
\end{align}
Since the massless fermions consists of singlets and vectorlike pairs, there is no further gauge breaking.

\subsection{Sp(6)$+\ydiagram{1}+\ydiagram{1,1,1}$}
We have
\begin{align}
    \text{Sym}^2(\mathbf{6}) =& \mathbf{21}
    \\
    \text{Sym}^2(\mathbf{14}') =& \mathbf{21}+\mathbf{84}
    \\
    \mathbf{6}\otimes \mathbf{14}' =& \mathbf{14}+\mathbf{70}
\end{align}
The adjoint $\mathbf{21}$ always leaves a residual $U(1)$. For the remaining bilinear channels we can decompose under the maximal subgroup
\begin{align}
    SU(2)\times U(1)\times U(1)\subset& \text{ Sp}(6),
\end{align}
to find
\begin{align}
    \mathbf{14}\downarrow_{SU(2)\times U(1)\times U(1)} =& \mathbf{1}_{2,0}
+\mathbf{1}_{0,2}
+2\mathbf{1}_{0,0}
+\mathbf{1}_{0,-2}
+\mathbf{1}_{-2,0}
+\cdots ,
    \\
    \mathbf{70}\downarrow_{SU(2)\times U(1)\times U(1)} =& \mathbf{1}_{2,2}
+2\,\mathbf{1}_{2,0}
+\mathbf{1}_{2,-2}
+2\,\mathbf{1}_{0,2}
+3\,\mathbf{1}_{0,0}
+2\,\mathbf{1}_{0,-2}
+\mathbf{1}_{-2,2}
+2\,\mathbf{1}_{-2,0}
+\mathbf{1}_{-2,-2}
+\cdots, 
    \\
    \mathbf{84}\downarrow_{SU(2)\times U(1)\times U(1)} =& \mathbf{1}_{2,2}
+\mathbf{1}_{2,0}
+\mathbf{1}_{2,-2}
+\mathbf{1}_{0,2}
+2\,\mathbf{1}_{0,0}
+\mathbf{1}_{0,-2}
+\mathbf{1}_{-2,2}
+\mathbf{1}_{-2,0}
+\mathbf{1}_{-2,-2}
+\cdots.
\end{align}
All of these provide a way to give a VEV that breaks just to $SU(2)$, but in particular we can do it with the NMAC $\mathbf{14}$, while the $\mathbf{70}$ and $\mathbf{80}$ are repulsive.

\bibliographystyle{JHEP}
\bibliography{references}
\end{document}